\documentclass{aastex}
\usepackage{aas_macros}
\usepackage{spr-astr-addons}
\usepackage{url}
\usepackage[latin1]{inputenc}
\usepackage[french, english]{babel}  
\usepackage{lscape}  
\usepackage{threeparttable} 
\usepackage{dcolumn,amsmath}
\usepackage{graphicx}
\usepackage{bm}
\usepackage{amsfonts}
\usepackage{amsthm}
\usepackage{epsfig}
\usepackage{fancyhdr}
\usepackage{minitoc}
\usepackage{natbib}
\usepackage{multicol}
\usepackage{longtable}
\usepackage[T1]{fontenc}

\RequirePackage{color}
\begin{document}
\title{Distributions of Pseudo-Redshifts and Durations (Observed and Intrinsic) of Fermi GRBs}
\author{H.~Zitouni}
\affil{ PTEA laboratory, Facult\'e des sciences, Universit\'e Dr
Yahia Fares, P\^{o}le urbain, M\'ed\'ea, Algeria.}
\email{zitouni.hannachi@univ-medea.dz} \and
\author{N.~Guessoum} \affil{Department of Physics, College of Arts
\& Sciences, American University of Sharjah, UAE.}\email{nguessoum@aus.edu}
\and
\author{K.~M.~AlQassimi} \affil{College of Arts \& Sciences, American University of Sharjah, UAE.}\email{b00065464@aus.edu}
\and
\author{O.~Alaryani} \affil{Physics Department, McGill University, Montreal, QC, H3A 2T8, Canada}\email{omar.alaryani@physics.mcgill.ca}
\begin{abstract}
Ever since the insightful analysis of the durations of gamma-ray
bursts (GRBs) by \cite{Kouveliotou:93}, GRBs have most often been
classified into two populations: ``short bursts" (shorter than 2.0
seconds) and ``long bursts" (longer than 2.0 seconds). However,
recent works have suggested the existence of an intermediate
population in the bursts observed by the Swift satellite.
Moreover, some researchers have questioned the universality of the
2.0-second dividing line between short and long bursts: some
bursts may be short but actually result from collapsars, the
physical mechanism behind normally long bursts, and some long ones
may originate from mergers, the usual progenitors of short GRBs.

In this work, we focus on GRBs detected by the Fermi satellite
(which has a much higher detection rate than Swift and other
burst-detecting satellites) and study the distribution of their
durations measured in the observer's reference frame and, for
those with known redshifts, in the bursts' reference frames.
However, there are relatively few
bursts with measured redshifts, and this makes an accurate
study difficult. To overcome this problem, we follow
\cite{Zhang:2018} and determine a ``pseudo-redshift" from the
correlation relation between the luminosity $L_p$ and the energy
$E_p$, both of which are calculated at the peak of the flux.
Interestingly, we find that the uncertainties in the quantities
observed and used in the determination of pseudo-redshifts, do
affect the precision of the individual results significantly, but
they keep the distribution of pseudo-redshifts very similar to
that of the actual ones and thus allow us to use pseudo-redshifts
for our statistical study. We briefly present the advantages and
disadvantages of using pseudo-redshifts in this context.

We use the reduced chi-square and the maximization of the
log-likelihood to statistically analyze  the distribution of Fermi
GRB durations. Both methods show that the distribution of the
observed (measured) and the intrinsic (source/rest frame) bursts
durations are better represented by two groups/populations, rather
than three.

\end{abstract}

\keywords{Gamma-rays: bursts, theory, observations; Methods: data
Analysis, statistical, chi-square test, likelihood.}

\section{Introduction}
Gamma-ray bursts (GRBs) are the most energetic electromagnetic
events known in the cosmos. They have been studied in depth since
their discovery in the late 1960's (first reported by
\cite{Klebesadel:1973}). While the cosmological (extra-galactic)
origin of the GRBs is well established (since it was confirmed by
the spatially isotropic distribution of bursts detected by the
CGRO/BATSE instrument in the 1990's and from the redshift
measurements, starting with \cite{Metzger:1997}), questions about
their sources and their physical nature remain incompletely
resolved. It is, however, commonly assumed that GRBs result from
different, heterogeneous populations and characteristics (see,
most recently,
\citet{{Zhang:2014},{Shahmoradi:2015},{Chattopadhyay:2017}}).

Various descriptive and statistical studies have been conducted
using GRB data obtained with instruments that work in several
bands of the electromagnetic spectrum.
The duration of a burst is one of
the most commonly used parameters for classifying and
characterizing gamma-ray bursts (from \cite{Kouveliotou:93} to
\citet{{Zitouni:15},
{Tarnopolski:2015},{Tarnopolski:2016b},{Tarnopolski:2016a},{Tarnopolski:2016c},{Kulkarni:2017},
{Zhang:2018}}). Burst durations are most often defined by
$T_{90}$, the time interval over which 90\% of the total fluence
(integrated flux, minus the background) is recorded
\citep{{Kouveliotou:93}, {Koshut:1995}}

Most of the statistical studies on the distribution of $T_{90}$
durations are consistent with the existence of two populations of
bursts, but a relatively small number of works have found in the
data indications that a third, intermediate group might exist
(\citet{{Horvath:1998},{Balastegui:2001},{Chattopadhyay:2007},{Horvath:2008},{Zitouni:15},{Zhang:2016MNRAS},{Kulkarni:2017}}).
The existence of two different populations is now fully confirmed:
a consensus exists on the hypernova-collapsar nature of long
bursts, and with the August 2017 observational confirmation of a
merger of two neutron stars (via the detection of both
gravitational waves and electromagnetic radiation from a short
GRB, GW/GRB170817 -- \citet{{Abbott2017b},{Abbott2017a}}), the
merger nature of short bursts appears to now be firmly
established. Still, a third (intermediate) population is not
totally rejected.

In this work, we wish to explore the distribution of Fermi GRB
durations, using the data published on its
website\footnote{https://heasarc.gsfc.nasa.gov/W3Browse/fermi}
\citep{vonKienlin2014, Bhat:2016ApJS}.
In particular, we seek to investigate the {\it intrinsic} durations,
$T_{90}^{r}$, i.e. as they would appear in
the bursts' own source frames, that is before the durations are
dilated due to the cosmological expansion effects. Such an
investigation requires the knowledge of both the observer-measured
durations and the redshifts of the bursts.

The Fermi GRB database contains more than 2,200 events, of which
about 15\% are short bursts ($T^{Obs}_{90} < 2.0$ s). They all
have observer-measured durations, but only about 6\% of them have
redshifts obtained by measurements. To overcome this limitation,
we use the correlation relation between the luminosity at the peak
of the photon flux, $L_p$, and the photon energy at the peak,
$E_p$, to determine a ``pseudo-redshift". For this, we have two
correlation relations: one for long bursts obtained by
\citet{{Yonetoku:2004},{Yonetoku:2010}}, and the other for short
bursts obtained by \cite{Tsutsui:2013} and \cite{Zhang:2018}. We
validate this procedure of obtaining reasonably correct
``redshifts" by comparing the pseudo-redshifts obtained from the
correlation relation with those that are available from
measurements. In the process, we
 note that the substantial uncertainties over $L_p$ and $E_p$
 have a significant impact on individual pseudo-redshift values but
do not affect the overall
redshift distributions. This allows us to confidently use
pseudo-redshifts to study the distribution of intrinsic durations
of bursts.

\section{Study of observed distributions}
We must take a number of necessary precautions in the selection of
our burst sample, e.g. that the photon flux be above the Fermi threshold
 ($P_{th}\simeq\,0.75$ photons cm$^{-2}$ s$^{-1}$), and that the
energy $E_p$ be in the Fermi-detection interval [8
keV, 1000 keV] \citep{Bhat:2016ApJS}. Additionally, bursts that lack
some data or have quantities given with large uncertainties
must be eliminated. This precaution will be explained later.

\subsection{Distribution of observed durations of Fermi GRBs}
\begin{figure}
\begin{tabular}{c}
   \includegraphics[angle=0, width=0.45\textwidth]{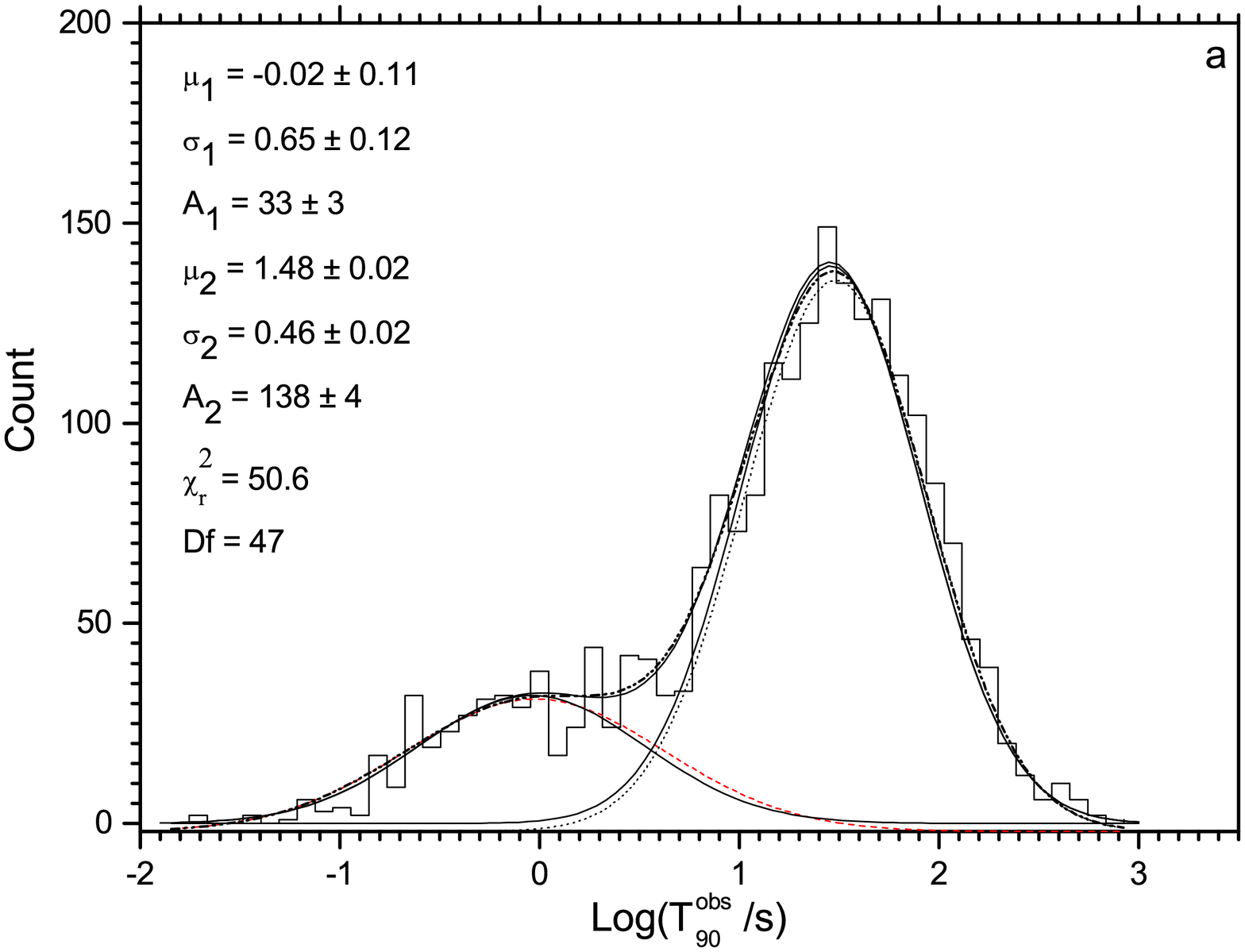} \\
   \includegraphics[angle=0, width=0.45\textwidth]{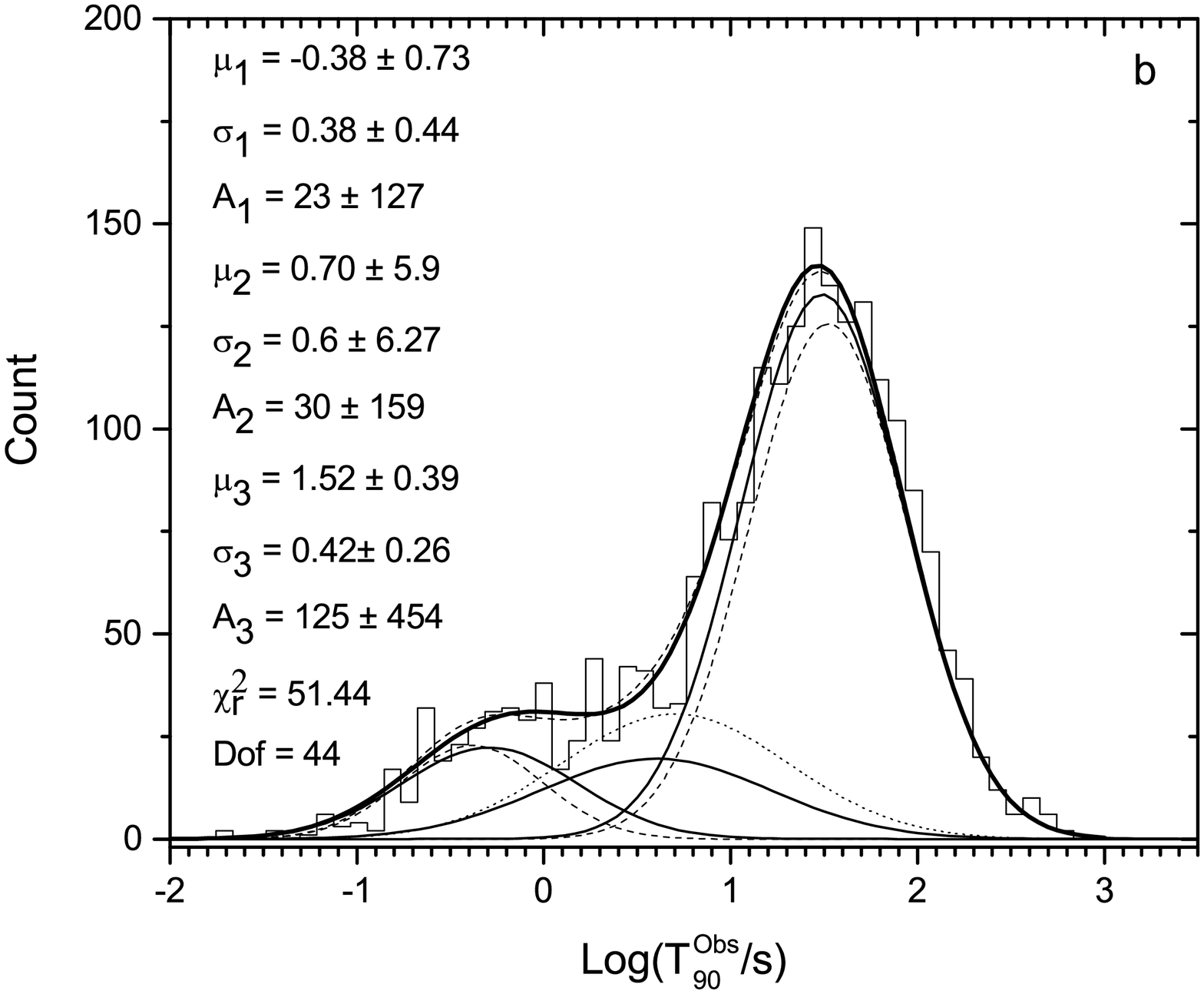}
\end{tabular}
\caption{Observed Duration distribution of 2238 Fermi's GRBs. The
parameters  A, $\mu$, and $\sigma$ characterize the Gaussian fit.
N=2238 GRBs and the bin width is 0.09. The dotted lines correspond
to the results of the $\chi^2_r$ method and the continuous lines
correspond to the results obtained by log-likelihood method.}
 \label{fig1N}
\end{figure}

There are 2239 bursts observed by Fermi/GBM: 368 short ones
($T^{Obs}_{90} < 2.0~s$) and 1870 long ones ($T^{Obs}_{90} >
2.0~s$) up to 13 January 2018 (a single burst, GRB 090626707, has
no reported duration). The distribution of the observed durations
of this sample (of 2238 GRBs) is shown in Figure (\ref{fig1N}). We
fit this data/distribution using two methods: by minimization of
the reduced chi-square function, $\chi^2_r$, and by maximization
of the log-likelihood function. The reduced chi-square function is
defined by:
\begin{equation}
   \chi^2_r=\frac{\sum_i^n (O_i-E_i)^2}{Dof},
\end{equation}
where n is the total number of bins; $O_i$ and $E_i$ are,
respectively, the observed and expected numbers of bursts in the i-th
bin; Dof is the number of degrees of freedom, $ Dof = n-k$, where
k is the number of independent fitting parameters
\citep{Andrae:2010}. The results of the $\chi^2_r$ minimization
method are presented in Figure (\ref{fig1N}); the parameters of
each fit are given in the two corresponding sub-figures. Based on
the $\chi^2_r$ values, fits using two groups/populations and three
groups/populations are equally good.

The second statistical method (maximizing the log-likelihood
function) is based on the
 function $L(x;\theta)$, defined by:
\begin{equation}
   L(x;\theta)=\prod_i^n f(x_i;\theta) ,
\end{equation}
where $f(x_i;\theta)$ is the probability density
associated with the measured data $x_i$, and $\theta$ denotes the
parameters for the model. This method finds
the values of the model parameters $\theta$ that maximize the
likelihood function $L(x;\theta)$.  Simply put, this selects the
parameter values that make the data most probable. In practice, it
is often convenient to work with the natural logarithm of the
likelihood function, thus referred to as ``log-likelihood", $\ln{L}$.

Using both methods, we have
studied the distribution of durations by representing it with
weighted sum of two or three Gaussian functions:
\begin{equation}
   f(\log{T_{90}^{Obs}}) = \sum_{i}^{n}~\frac{f_i}{\sigma_i\sqrt{2\pi}}~exp(-\frac{1}{2}(\frac{\log{T_{90}^{Obs}}-\mu_i}{\sigma_i})^2),
\end{equation}
where the parameters $f_i$, $\mu_i$, and $\sigma_i$ characterize
the group $i$ from 1 to $n$, $n$ being the number of groups,
$f_i$ representing a weight of the
$i^{th}$ component or group, with $\sum\limits_{i=1}^n f_i=1$.

To calculate the number of bursts `Count(j)' in each channel j, of width
`binw', we use the following expression:
\begin{equation}
   Count(j) = \sum_{i}^{N}~A_i~exp(-\frac{1}{2}(\frac{\log{T_{90,j}^{Obs}}-\mu_i}{\sigma_i})^2),
\end{equation}
where the constant $A_i$  is calculated by: $N\times binw
\times\frac{f_i}{\sigma_i\sqrt{2\pi}}$, N being the total number of GRBs.

For the log-likelihood method, we have employed two information
criteria to estimate the quality of one model over another: the
Akaike Information Criterion (AIC) and the Bayesian Information
Criterion (BIC), which are estimators of the relative quality of
statistical models for a given set of data
\citep{{Akaike:1974ITAC},{schwarz:1978},{Kass:1995},{Kenneth:2004}}.
Thus, AIC and BIC provide means for model selection. They are defined by:
 \begin{eqnarray}\label{aic_bic}
   AIC &=& 2\,k - 2\ln{L},\\
   BIC &=& k\ln{N} - 2\ln{L},
 \end{eqnarray}

 The model with the lowest BIC or AIC is preferred. The results
of the maximum likelihood method are presented in Table
(\ref{tab1N}). Both the AIC and the BIC show that the
three-group model does not produce any improvement over the two-group
model. The parameters of each model are summarized in the two
corresponding sub-tables.

In Figure (\ref{fig1N}), we plot the curves corresponding to the
results obtained using the $\chi^2_r$ minimization method (dotted
lines) and the curves corresponding to the results obtained using
the log-likelihood method (solid lines). This result concords with
the recent works of \citet{{Tarnopolski:2015}} and \citet{{Kulkarni:2017}}.

Although the $\chi^2_r$ minimization method depends strongly on
the number of bins used \citep{{Huja:2009},{Tarnopolski:2015}}, it
does give a good fit, similar to what is obtained using the
log-likelihood function method. This is due to the large number of
data points, which exceeds 2000, and the appropriate density of
the data.

\begin{table}[ht]
\renewcommand{\arraystretch}{1.5}
  \centering
\caption{The distribution of the observed durations of 2238 Fermi GRBs
represented by two and three Gaussian functions (two and three GRB groups).
The parameters presented here are all defined in the main text.}
 \label{tab1N}
  \begin{tabular}{c|cc}
    \hline
    ~  & Two groups&  \\
    \hline
    i&1&2\\
    \hline
    f & $0.225\pm0.005$ & $0.775\pm0.005$ \\
    $\mu$ & $-0.04^{+0.22}_{-0.16}$ & $1.46^{+0.07}_{-0.04}$ \\
    $\sigma$ & $0.57^{+0.11}_{-0.03}$ & $0.45^{+0.02}_{-0.03}$\\
    $\ln{L}$&-2383.3&\\
    BIC&4805.2&\\
    AIC&4776.6&\\
    k&5&\\
    \hline
  \end{tabular}
  \begin{tabular}{c|ccc}
    \hline
    ~  & &Three groups&  \\
    \hline
    i&1&2&3\\
    \hline
    f & $0.13\pm0.05$& $0.15\pm0.13$&$0.72\pm0.07$\\
    $\mu$  & $-0.29\pm0.04$&$0.60^{+0.1}_{-0.4}$&$1.49\pm0.07$\\
    $\sigma$ &$0.47^{+0.17}_{-0.05}$&$0.62\pm0.13$&$0.44\pm0.01$\\
    $\ln{L}$&-2382.14&&\\
    BIC&4825.99&&\\
    AIC&4780.28&&\\
    k&8&&\\
    \hline
  \end{tabular}
  \end{table}

In a second step, we note that due to the uncertainties on $T_{90}^{Obs}$,
bursts near the value of 2.0 s might be
  short or long, and thus the number of short Fermi/GBM bursts may vary
  between 289 and 451, and the number of long bursts may vary between 1787 and 1949.
  We thus limit ourselves to the 1787 bursts that are most probably long and the 289 ones that are most probably short.
  We eliminated 162 bursts in total, of which 151 were originally ``short'', due to their large uncertainties, which made
  them potentially belong to either group.
This procedure of strongly separating the two groups is done in order to estimate
the intervals that cover the majority of the values of the spectra parameters
($\alpha$, $\beta$ and $E_p$) for each population, if we consider a two-group model.
We use those intervals in our subsequent calculations in cases where the values of the spectral parameters given by the NASA website are tainted with a large uncertainty.

\begin{figure}[h]
\centering
\includegraphics[angle=0, width=0.483\textwidth]{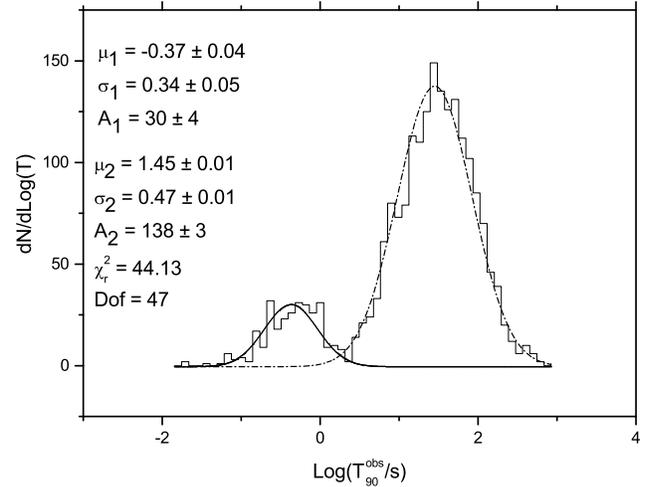}
\caption{Observed Fermi/GBM durations of selected 2076 Fermi GRBs.
The bin width is 0.09.} \label{fig3Nb}
\end{figure}

After applying the last filter, we plot in Figure (\ref{fig3Nb})
the distribution of durations of the remaining 2076 bursts. The
fits are again obtained using both the minimization of the
chi-square function and the maximization of the log-likelihood
function (Table \ref{tab2N}). A very slight difference is noted
between the results from the two methods. We note that after
removing the doubtful bursts, we obtain two populations that are
clearly separate.

\begin{table}[ht]
\renewcommand{\arraystretch}{1.5}
  \centering
\caption{Central value, $\mu$, and variance $\sigma^2$ of the
distribution of 2076's Fermi duration. Here we use the same
definitions as in Table (\ref{tab1N}).}
 \label{tab2N}
  \begin{tabular}{c|cc}
    \hline
    ~  & Two groups&  \\
    \hline
    i&1&2\\
    \hline
    f & 0.139 & 0.861 \\
    $\mu$ & -0.399 & 1.443 \\
    $\sigma$ & 0.377 & 0.453\\
    $\ln{L}$&-2041.50&\\
    BIC&4121.20&\\
    AIC&4093.0&\\
    k&5&\\
    \hline
  \end{tabular}
    \end{table}

\subsection{Fermi Distribution of $E_p$, $\alpha$ and
$\beta$}\label{subsec} \cite{Band:1993} assumed that the spectrum
of the gamma burst can be described by a function composed of two parts:

 $N(E)= N_0 \times$ \vspace{-0.3cm}
\begin{equation} \label{eq1}
\left\{
\begin{split}
 & (\frac{E}{100~keV})^{\alpha}\exp(-\frac{E}{E_0}) &, E\leq E_b
    \\
 & (\frac{E_b}{100~keV})^{\alpha-\beta}\exp{[\beta-\alpha](\frac{E}{100~keV})^{\beta}}&,
E>E_b
\end{split}
\right\}.
\end{equation}
where $E_b = (\alpha - \beta)E_0$ and $E_0=\frac{E_p^{Obs}}{\alpha+2}$. $N(E)$
has units of photons $s^{-1}$ $keV^{-1}$ $cm^{-2}$.

\begin{figure}
\begin{tabular}{c}
   \includegraphics[angle=0, width=0.45\textwidth]{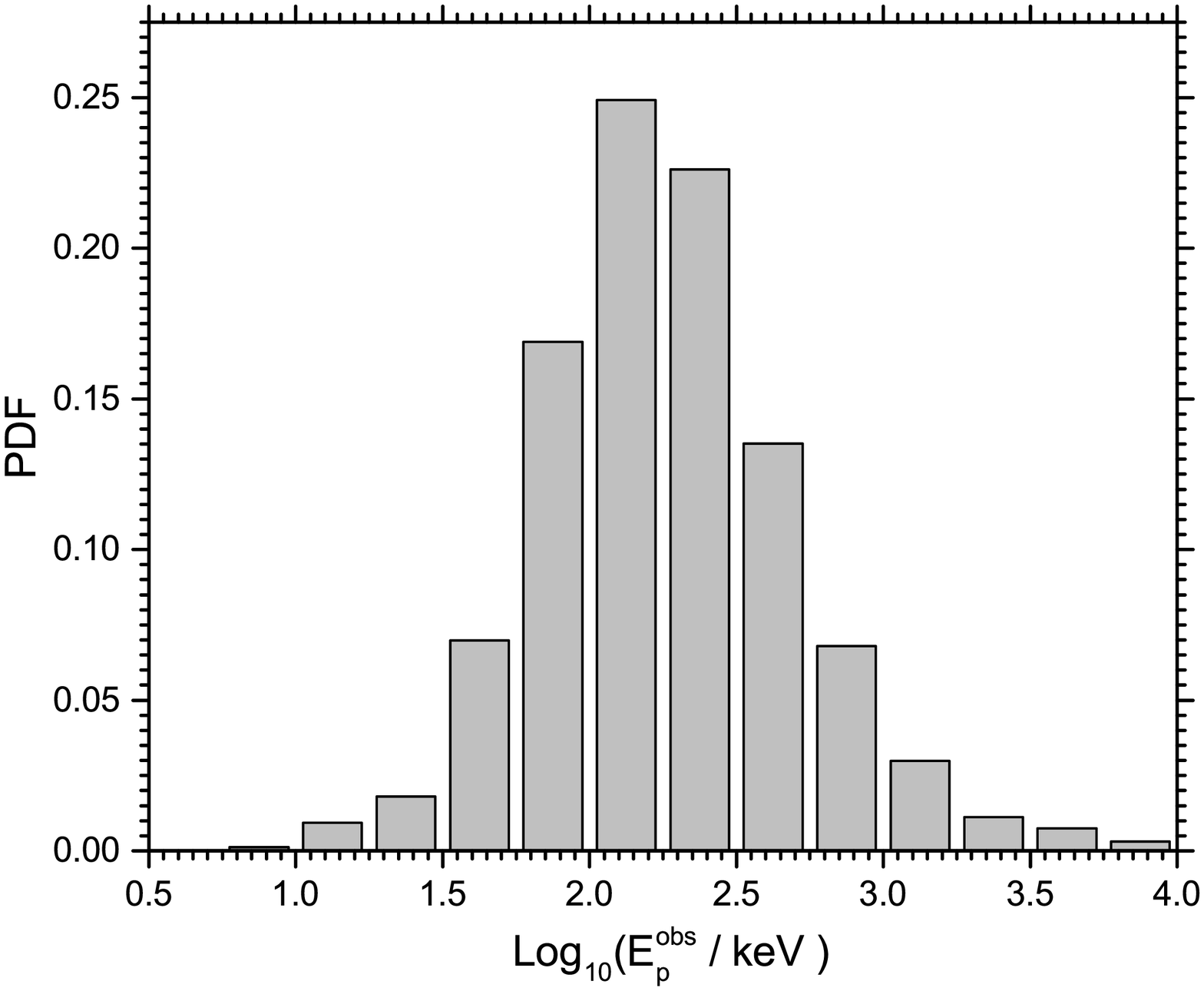} \\
   \includegraphics[angle=0, width=0.45\textwidth]{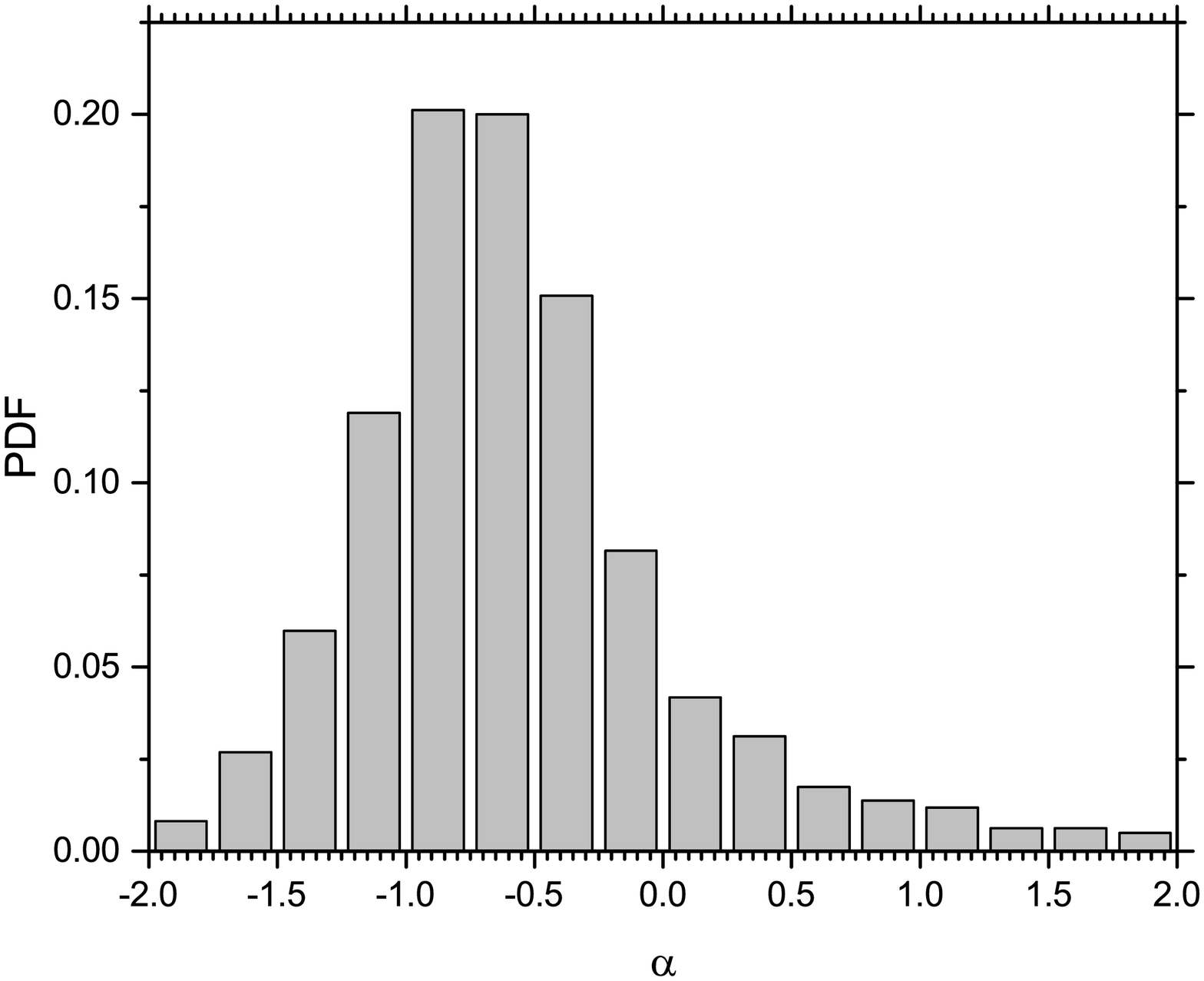}\\
   \includegraphics[angle=0, width=0.45\textwidth]{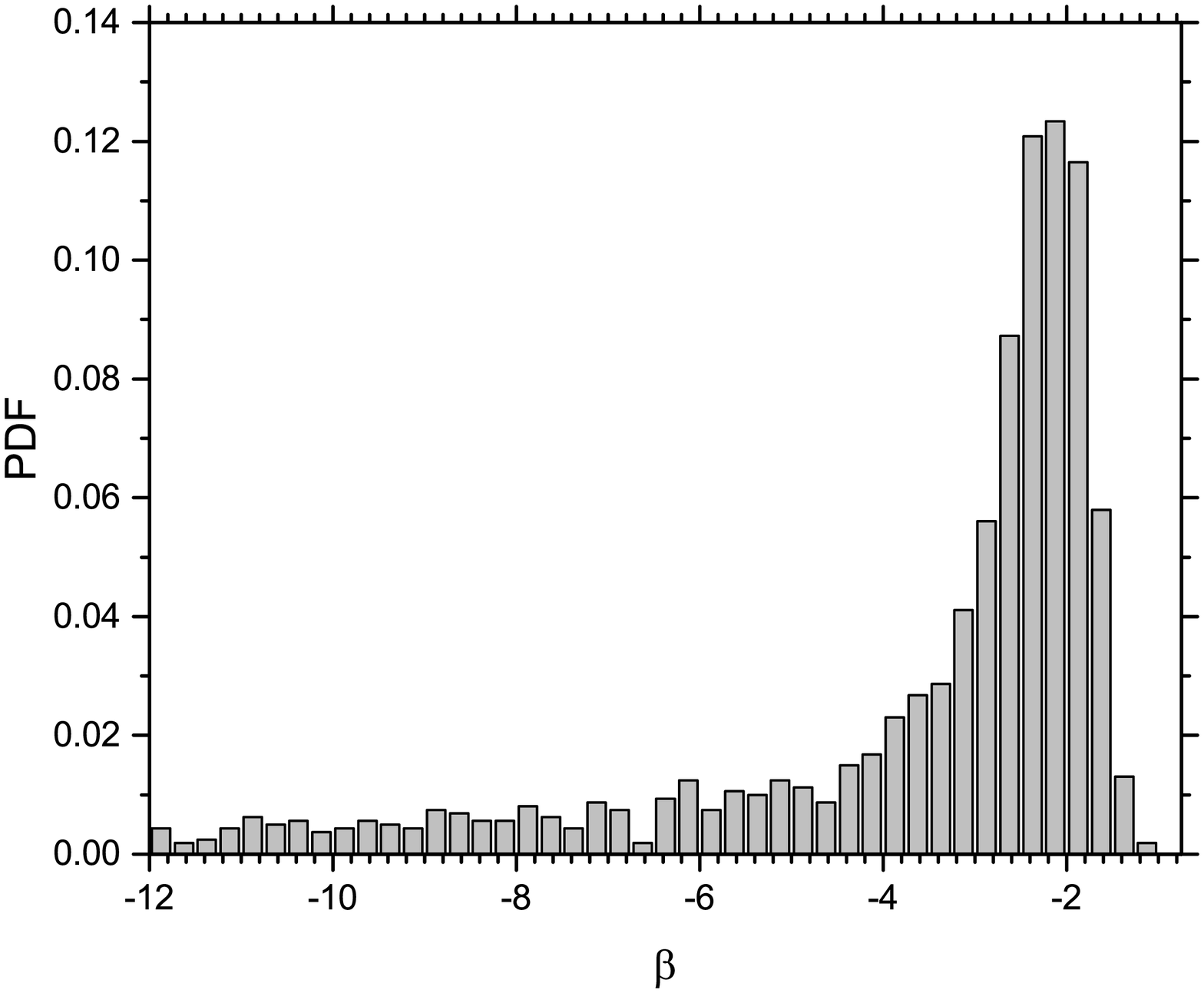}
\end{tabular}
\caption{Distributions of $E_p^{obs}$, $\alpha$ and $\beta$ for a
sample composed of 1605 Fermi LGRBs.} \label{fig456}
\end{figure}

\begin{figure}

\begin{tabular}{c}
   \includegraphics[angle=0, width=0.45\textwidth]{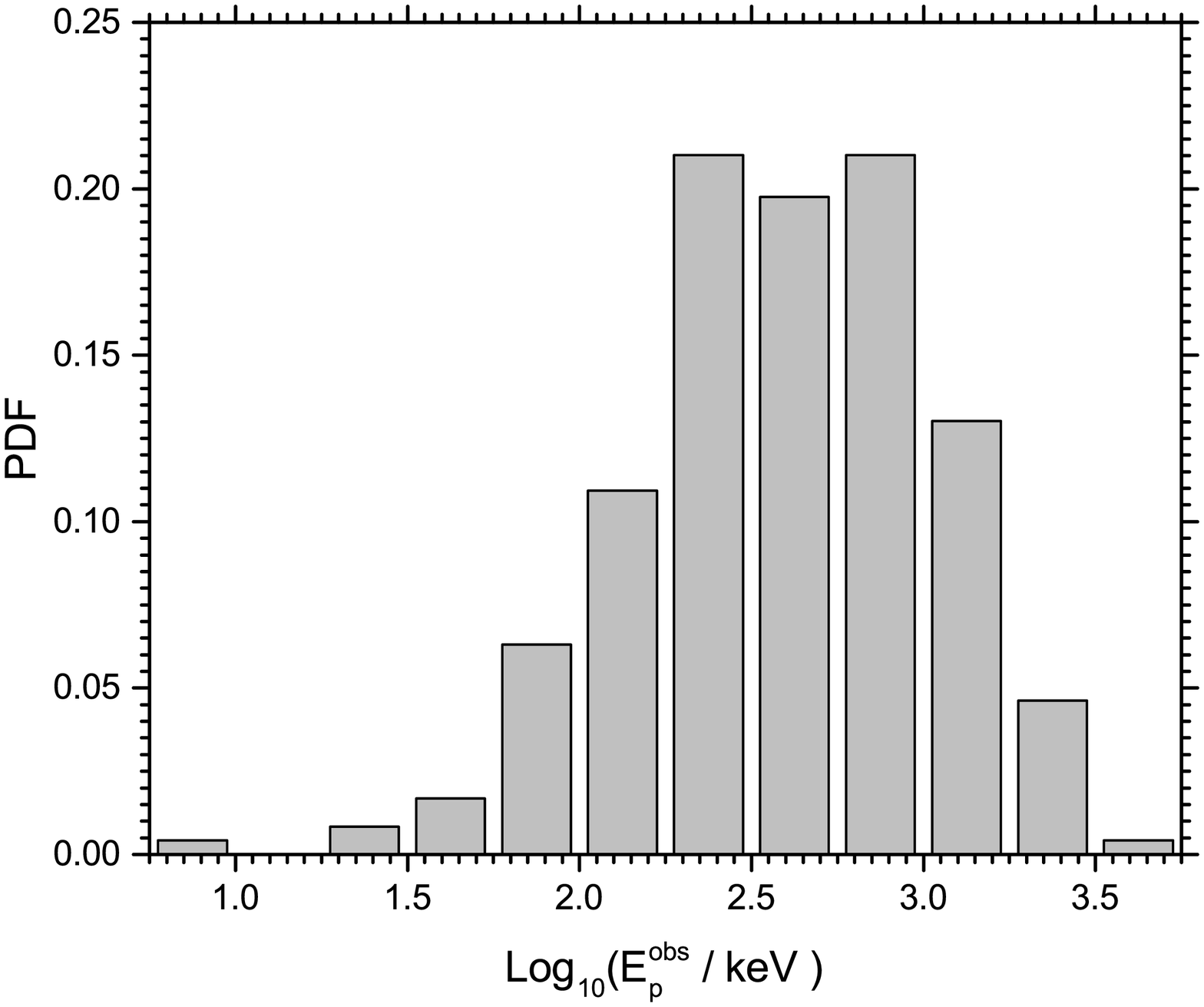} \\
   \includegraphics[angle=0, width=0.45\textwidth]{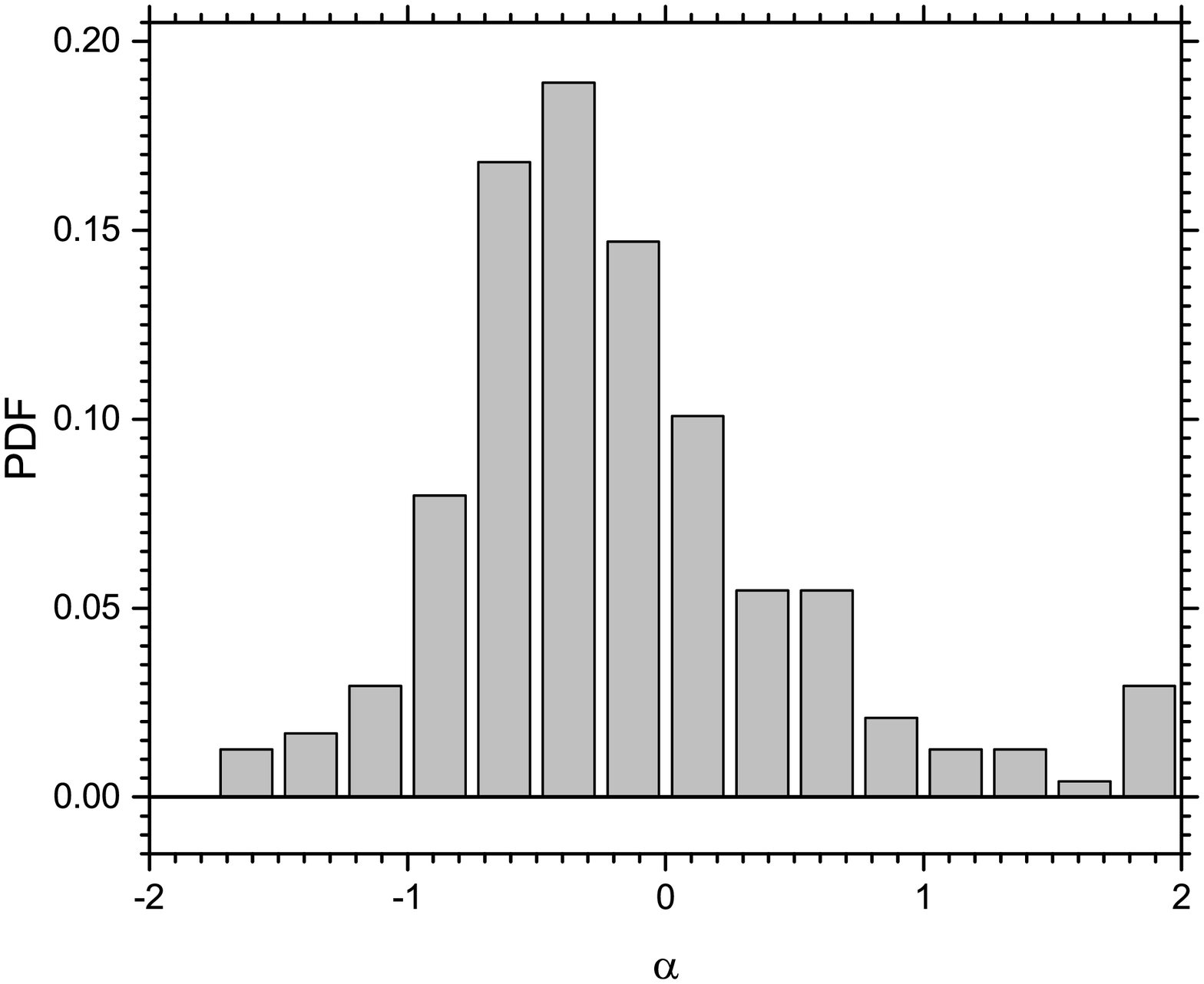}\\
   \includegraphics[angle=0, width=0.45\textwidth]{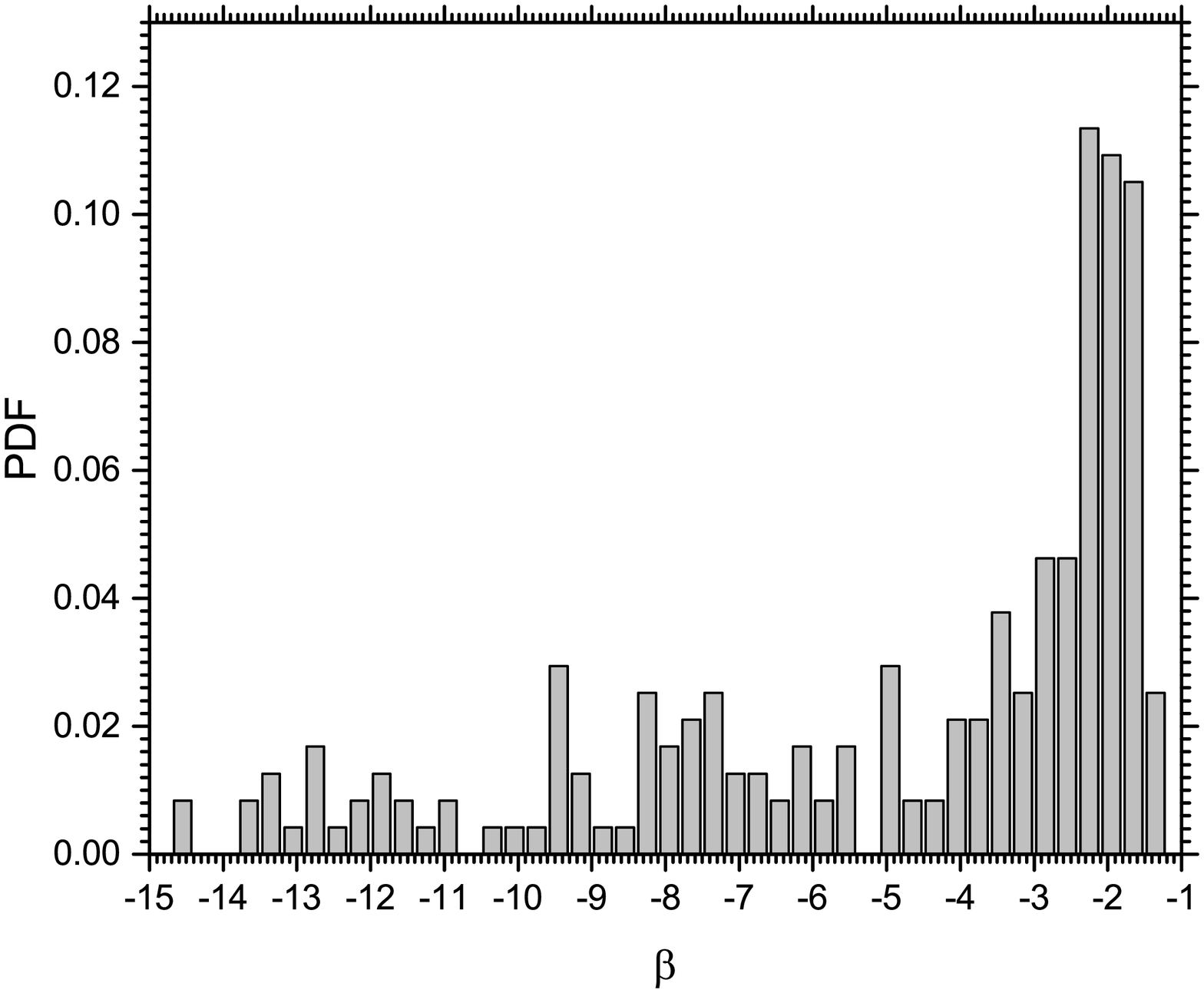}
\end{tabular}
\caption{Distributions of $E_p^{obs}$, $\alpha$ and $\beta$ for a
sample composed of 238 Fermi SGRBs.} \label{fig789}
\end{figure}

We seek the values of $E_p^{obs}$, $\alpha$ and $\beta$
for the two samples of gamma-ray bursts that we selected above
(those consisting of 289 short SGRBs and 1787 long LGRBs). Out of
the 289 SGRBs, we find 238 with definite spectral parameters. Out
of the 1787 LGRBs, 1605 have definite spectral parameters. We
present the distributions of these quantities (as probability density functions, PDF)
in Figures
(\ref{fig456}) and (\ref{fig789}). In these graphical
representations, we apply no filter.

We note that for the two types of bursts, the values of $\alpha$
generally fall in the range $(-2,2)$, while the values of $\beta$
are less than $-1$. The lowest values of $\beta$ can go beyond
$-10$ ($20\%$ of the  238 SGRBs and $30\%$ of the 1605 LGRBs). The
distributions of $E_p$ values as presented graphically in Figures
(\ref{fig456}) and (\ref{fig789}) show that around $95\%$ of the
bursts have $E_p$ energies in the range $[8,1000]$ keV.

In the calculations that we subsequently perform, we consider
values for $\alpha$, $\beta$, and $E_p$ as shown in Table
(\ref{distrib}) for all groups/populations (short, intermediate,
or long). These intervals have been adopted by taking into account
the figures (\ref{fig456}) and (\ref{fig789}) as well as the
limits from BATSE given in the catalog 5B
\citep{Goldstein:2013zua} so that the resulting distributions are
centered around the modes; for $E_p$ we take the Fermi detection
range. The choice of our intervals, which are very close to those
presented in the data of the fifth catalog of BATSE cover the
majority of the bursts.

\begin{table}[ht]
  \centering
\caption{Ranges of the spectral parameters $\alpha$, $\beta$, and
$E_p$, with their uncertainties.}\label{distrib}

  \begin{tabular}{|r|c|}
    \hline
    parameter & Interval\\
    \hline
    $\alpha\pm\Delta\alpha$ & [-2.0 ,  1.0]\\
    $\beta\pm\Delta\beta$ & [-4.5 , -1.0]\\
    $E_p\pm\Delta E_p (keV)$ & [8 , 1000]\\
    \hline
  \end{tabular}
  \end{table}

\section{Determination of Pseudo-Redshifts}
As mentioned above, to address the dearth of redshifts and give
ourselves a way to investigate intrinsic burst durations, we use
correlation relations to determine pseudo-redshifts. This idea has
been proposed and used several times in the past (e.g.
\citep{{Atteia:2003},{Kocevski:2003},{Ghirlanda:2005MNRAS},{Guidorzi:2005MNRAS},{Guidorzi:2006MNRAS},{Tsutsui:2008AIPC},{Tsutsui:2013},{Zhang:2018}}).
We delineate the method  succinctly in the following sub-section.

\subsection{Can correlation relations give correct redshifts?}
\label{section}

We use a sample of Fermi bursts with known redshifts to assess the
extent to which the correlation relations of
\cite{{Yonetoku:2004},{Yonetoku:2010}} for long bursts and that of
\cite{Tsutsui:2013} for short bursts can produce correct
redshifts.

The luminosity at the peak of the flux can be calculated using:
\begin{equation}{\label{eq4N}}
\frac{L_p}{erg~s^{-1}} = 4\pi D_L^2(z)~F_{\gamma}~k_c,
 \end{equation}
where $L_{p}$ is k-corrected with the method developed by
 \cite{Bloom:2001}, and $k_c$ is the proper k-correction factor
 \citep{{Yonetoku:2004},{Rossi:2008},{Elliott:2012},{Zitouni:2014}} defined
 by:
 \begin{equation}\label{eq_kc}
k_c= \frac{\int_{E_{1}/(1+z)}^{E_{2}/(1+z)}E N(E)
dE.}{\int_{E_{min}}^{E_{max}}E N(E) dE.},
 \end{equation}
where $[E_1=1\,keV;E_2=10^4\,keV]$ is the gamma radiation band in
the source's frame and $[E_{min}=8keV; E_{max}=10^3keV]$ is
Fermi-detection interval.

 The peak energy flux, denoted by $F_{\gamma}$ and calculated in
$\mathrm{erg\,cm^{-2}\,s^{-1}}$, is calculated numerically through
the following equation:
\begin{equation}\label{eq333}
F_{\gamma}= \int_{E_{min}}^{E_{max}}E N(E) dE.
 \end{equation}

 $D_L(z)$ is the luminosity distance, which is expressed by the following equation:
 \begin{equation}
  D_L=\frac{(1+z)c}{H_0}\int_0^z
  \frac{dz'}{\sqrt{\Omega_M(1+z')^3+\Omega_{\Lambda}}}.
 \end{equation}
 $D_L(z)$ depends on the cosmological parameters $H_0$, $\Omega_m$, and $\Omega_{\Lambda}$.
We consider a flat universe, where the values of these parameters
are $70~km\,s^{-1}\,Mpc^{-1}$, 0.3, and 0.7.

For long bursts, the Yonetoku relation \citep{Yonetoku:2010} that we use is:
\begin{equation}{\label{eq7N}}
 \frac{L_p}{erg~s^{-1}} = 10^{52.97}~[\frac{E_p^{obs}(1+z)}{774.5~keV}]^{1.60},
 \end{equation}
where $E_p^{obs}$ is the photon energy at the peak of the spectrum, measured in keV,
and $E_p=E_p^{obs}(1+z)$ is the peak energy in the source's frame.

For short bursts, the correlation relation \citep{Tsutsui:2013} that we use is (in the source frame):
\begin{equation}{\label{eq8N}}
 \frac{L_p}{erg~s^{-1}} =
 10^{52.29\pm0.07}~[\frac{E_p^{obs}(1+z)}{774.5~keV}]^{1.59\pm0.11}.
 \end{equation}

In order to validate these correlation relations and their
capacity to determine redshifts, we use a sample of 117 Fermi
bursts with known redshifts, which we give in Table (\ref{tab3}).
A systematic and thorough search for redshifts of Fermi GRBs,
including  GCN telegrams, gave us 134 cases. This will surely
increase in the future. The 134 GRBs were reduced to 117 bursts
because the energy $E_p^{Obs}$ was sometimes outside of the Fermi
detectors' range $[8, 1000]$ keV. The bursts that we eliminated
are listed in Table (\ref{tab2}). \setcounter{table}{4}
\begin{table}[ht]
  \centering
\caption{17 GRBs eliminated due to their $E_p^{Obs}$ falling
outside of the Fermi detectors' range $[8, 1000]$ keV.}
\label{tab2}
  \begin{tabular}{ccccc}
    \hline
    Name  & z&   $E_b^{Obs}$\\
    & & keV\\
    \hline
    GRB160623209&0.367&1032.9\\
    GRB140809133&0.041&1078.76\\
    GRB130215063&0.579&1210.92\\
    GRB170604603&1.329&1285.12\\
    GRB120711115&1.4&1357\\
    GRB090902462&1.82&2152\\
    GRB150120123&0.46&$<$10\\
    GRB101213451&0.414&\\
    GRB171010792&0.3285&\\
    GRB090519881&3.85&\\
    GRB171222684&2.409&\\
    GRB100413732&4.0&\\
    GRB080905705&2.374&\\
    GRB120712571&4.1745&\\
    GRB170903534&0.886&\\
    GRB090510016&0.903&\\
    GRB171020813&1.87&\\
    \hline
  \end{tabular}

  \end{table}

\begin{figure}

\begin{tabular}{c}
   \includegraphics[angle=0, width=0.45\textwidth]{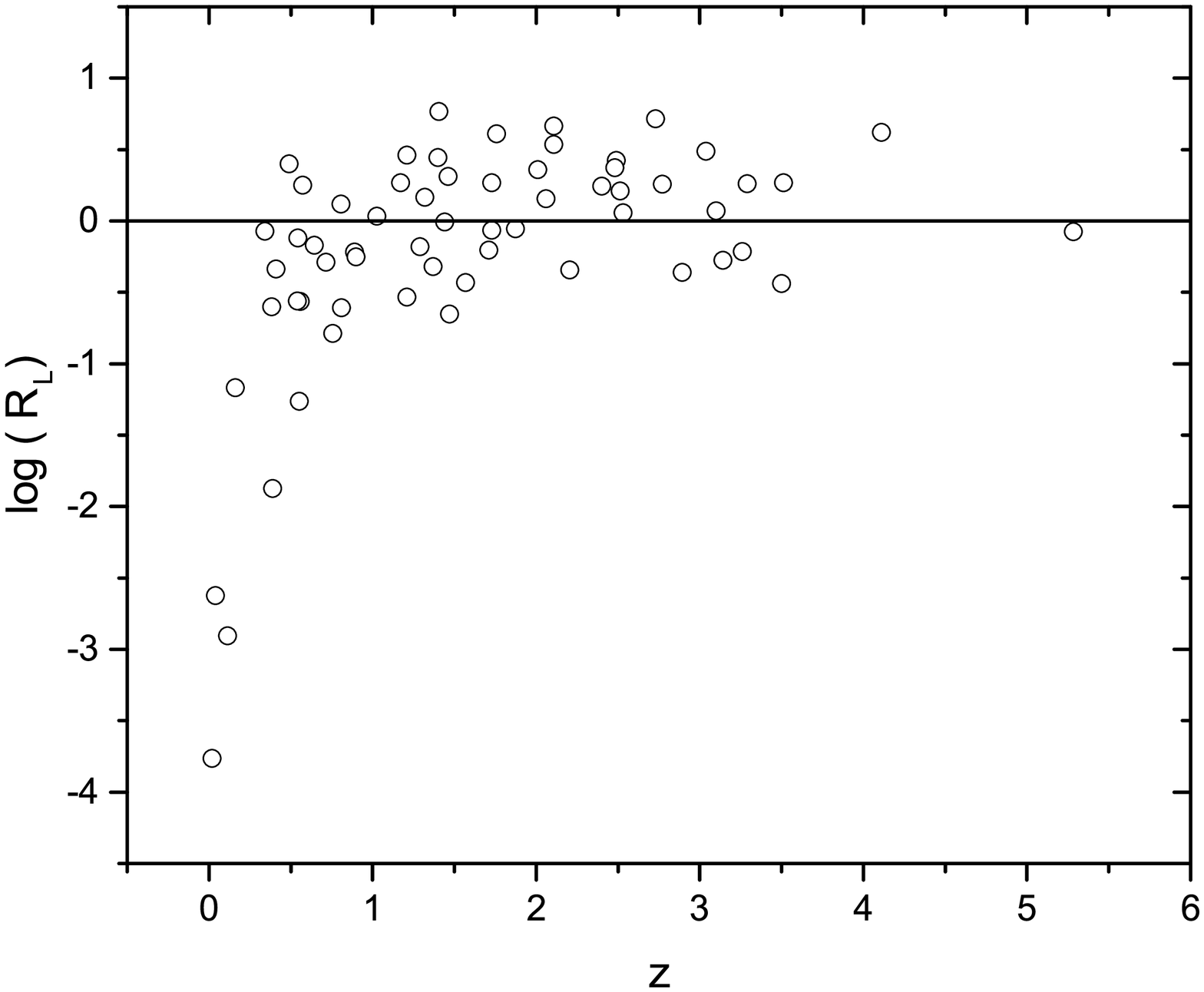} \\
   \includegraphics[angle=0, width=0.45\textwidth]{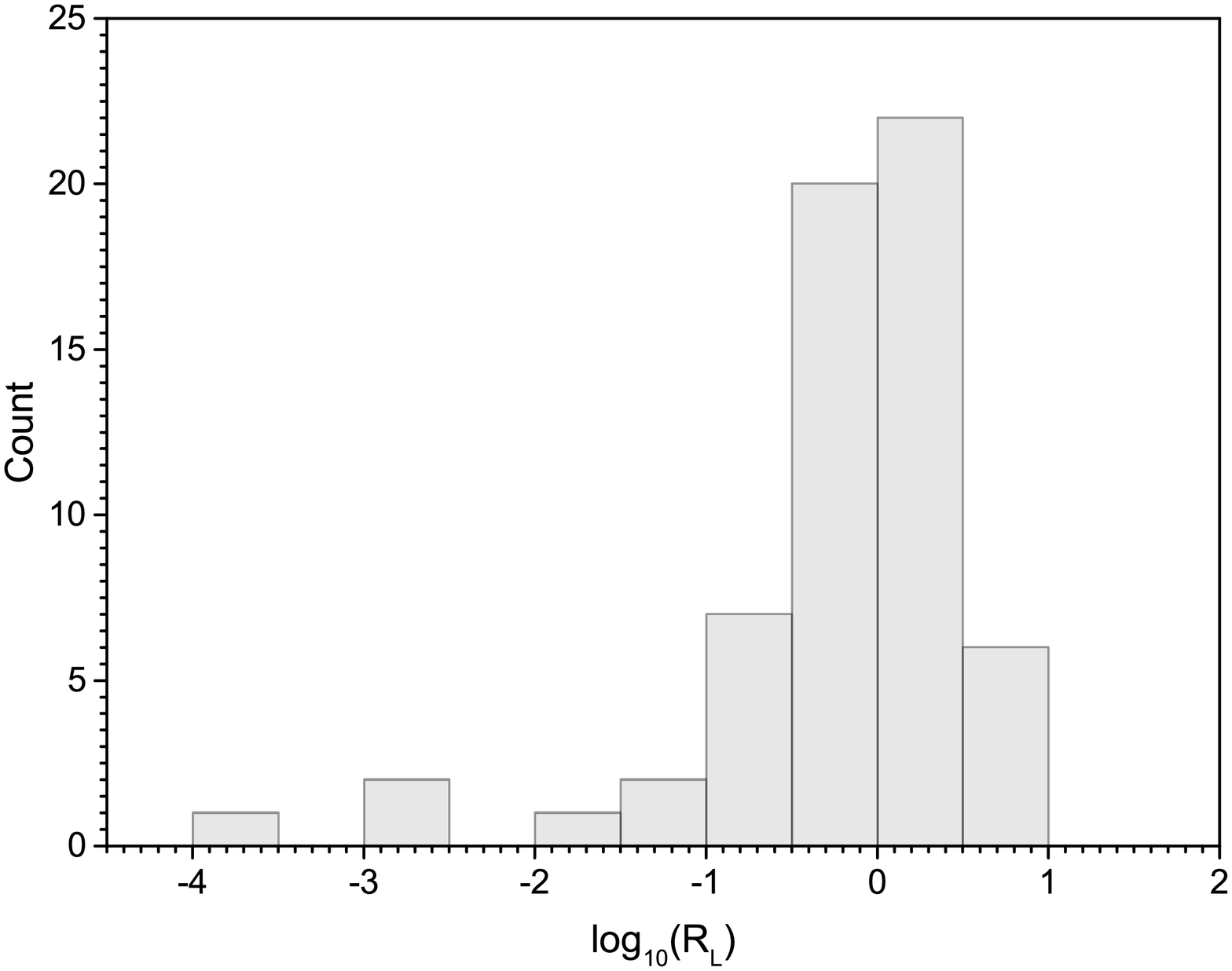}
\end{tabular}
\caption{Top: $\log{R_L}$ (defined in the text) as a function of
the redshift z for 61 GRBs. Bottom: Histogram of $\log{R_L}$. The
values of $\log{R_L}$ are given in Table (\ref{tab3}). }
\label{fig10}
\end{figure}

We start by calculating the ratio of
the luminosity $L_p$, which is obtained by using the relation
\eqref{eq4N}, and the luminosity obtained using the correlation
relations \eqref{eq7N} or \eqref{eq8N}. This ratio, which we
denote by $R_L$, is plotted in Figure (\ref{fig10}) as a function of
the redshift z. This was obtained using the data for $E_p^{Obs}$,
$\alpha$, and $\beta$ found on the Fermi
website\footnote{https://heasarc.gsfc.nasa.gov/W3Browse/fermi/fermigbrst.html}.

When the ranges of the spectral parameters given in Table
(\ref{distrib}) are taken into account, our sample further reduces
to 61 GRBs.

The quantity $R_L$ represents the quality of the correlation
relation as a function of the redshift, the ideal situation being
for $R_L = 1.$ $R_L$ shows substantial dispersion in the data,
although the data does show a correlation between the luminosity
$L_p$ and the energy $E_p$ measured in the source frame. In Figure
(\ref{fig10}), we present $\log{R_L}$ vs. the redshift for the 61
GRBs (top panel), and a histogram of $\log{R_L}$ (bottom panel).
We note that 55 points out of 61 (i.e. almost  90\%) have an $ R_L
$ between 0.1 and 10. More than 67\% of bursts have an $ R_L $
between 0.3 and 3. The values of $\log{R_L}$ are given in Table
(\ref{tab3}).

\begin{figure}[h]
\centering
\includegraphics[angle=0, width=0.483\textwidth]{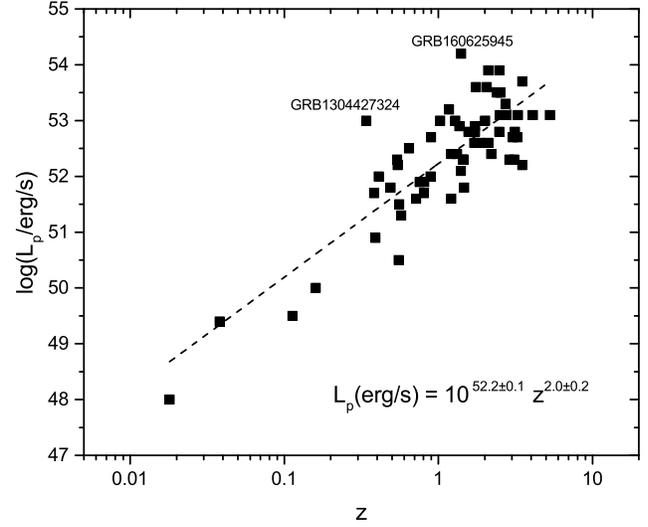}
\caption{$L_p$ versus z for 61 Fermi GRBs with known redshifts. The
values of $\log{R_L}$ are given in Table (\ref{tab3}).}
\label{fig12}
\end{figure}

It is also useful to study the evolution of the brightness in
terms of the redshift for the 61 GRBs. We present this evolution
graphically in Figure (\ref{fig12}). This can also be analytically
expressed by the following relation:
\begin{equation}\label{eq4}
  \frac{L_p}{10^{52.2\pm0.1}~erg\,s^{-1}} = z^{2.0\pm0.2}.
\end{equation}
The evolution of the luminosity $L_p$ in terms of the redshift $z$
is important to note. For this relation, there is no difference
between long and short bursts. The two bursts that are far from
the ``cloud'' are GRB130427324 and GRB160625945; they have the two
highest photon fluxes (523 $ph.\,cm^{-2}\,s^{-1}$ and 206
$ph.\,cm^{-2}\,s^{-1}$). It is then possible to use this relation
to infer pseudo-redshifts. In other references, e.g.
\citet{{Lloyd:2002},{Goldstein:2012}, {Salvaterra:2012},
{Geng:2013}, {Zhang:2014JApA}}, a correlation is sought between
$L_p$ and (1 + z) instead of z. We do not find the same slope at
low redshift, but our results are in agreement at high redshifts.
We do not focus on this issue, as it is not important in our
present subject.

As previously noted, our procedure for determining the
pseudo-redshift, noted $z_{cal}$, is based on solving the equation
\eqref{eq4N}=\eqref{eq7N} for long bursts and
\eqref{eq4N}=\eqref{eq8N} for short bursts. We have followed the
same approach as \cite{Zhang:2018}.

\begin{figure}[h]
\centering
\includegraphics[angle=0, width=0.483\textwidth]{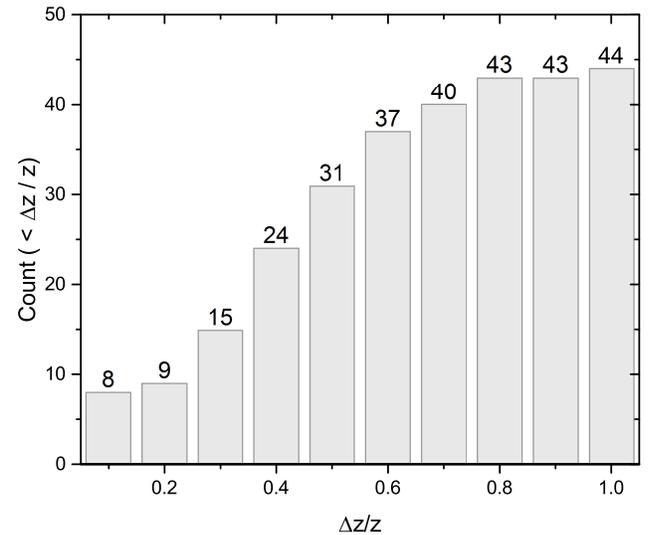}
\caption{$\frac{\Delta z}{z} =\frac{|z_{cal}-z_{obs}|}{z_{obs}}$.
`Count' represents the cumulative number of GRBs that have a
$\frac{\Delta z}{z}$ below the limit shown on the abscissa axis.
Here we have used the sample of 61 GRBs.} \label{fig13}
\end{figure}

\begin{figure}[h]
\centering
\includegraphics[angle=0, width=0.483\textwidth]{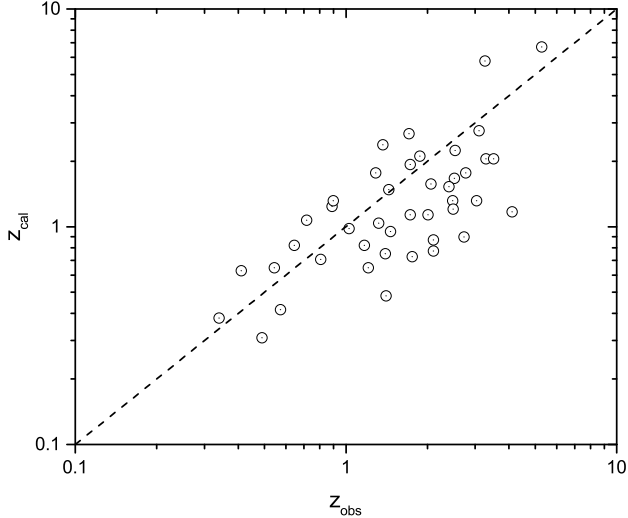}
\caption{$z_{cal}$ vs. $z_{obs}$ for the 43 GRBs with
$\frac{\Delta z}{z} < 0.8$. } \label{fig13plus}
\end{figure}

To assess our method for the determination of pseudo-redshifts, we
compare the values we determine ($z_{cal}$) with the measured
redshift values ($z_{obs}$) for the sample of 61 GRBs. For that,
we calculate the relative error, $\frac{\Delta z}{z}=\frac{\mid
z_{cal}-z_{obs}\mid}{z_{obs}}$, for each burst, then we represent
by a histogram (Fig.\ref{fig13}) the number of cases having a
relative uncertainty lower than a certain limit, which we vary
from 0 to 1. We note that for a relative uncertainty less than
0.5, we get 31 redshifts out of 61, i.e. $\sim$ 51 \%. Two-thirds
of the pseudo-redshifts have a relative uncertainty of less than
0.70. This rather large error fraction is due to the large
dispersion of the data ($\alpha$, $\beta$, $E_p$, and $P_{obs}$,
the photon flux). In Fig. (\ref{fig13plus}) we have plotted
$z_{cal}$ vs. $z_{obs}$ for the 43 GRBs that have a relative
uncertainty $\Delta z/z$ less than 80\%.

\subsection{Redshift determination of Fermi GRBs}
With the above validity tests of the method used to infer
pseudo-redshifts, we adopt our calculated values as {\it
statistically} valid substitutes for the actual redshifts $z$ for
 Fermi GRBs. We then take all the bursts recorded by the Fermi
satellite since (the first one on) 14 July 2008 (GRB080714086) and
up until 11 February 2018 (GRB180211754) \citep{vonKienlin2014,
Bhat:2016ApJS} without any imposed condition.

This new sample comprises 2266 GRBs: 378 short ones and 1888 long
ones. 243 GRBs lack one or more measured spectral parameter(s).
Taking into account the ranges given in Table (\ref{distrib}) for
the spectral parameters eliminates more than
 1000 GRBs.

\begin{figure}[h]
\centering
\includegraphics[angle=0, width=0.483\textwidth]{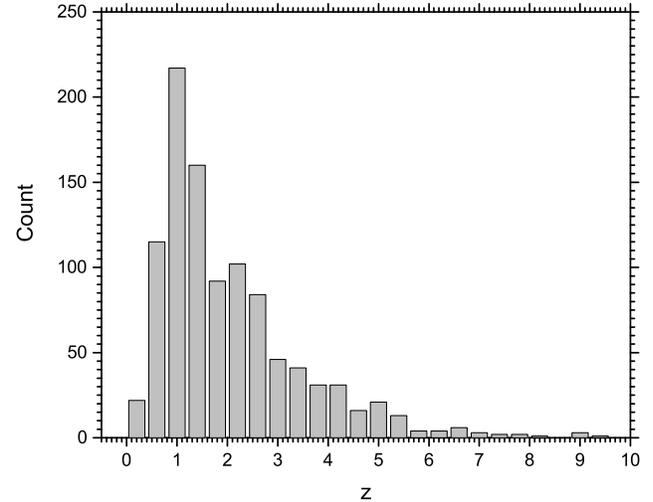}
\caption{Pseudo-redshift distribution for  1017 Fermi GRBs. The
bin width is  0.4.} \label{fig14}
\end{figure}

When trying to obtain pseudo-redshifts, in a few cases the
relevant equation has no solution, as the burst's parameters
($\alpha, \beta, E_p$) and the flux place it far from the
correlation line. In the end, we could determine pseudo-redshifts
for 1017 GRBs (144 SGRBs and 873 LGRBs). However, we stress that
while the values of the obtained pseudo-redshifts have large
uncertainties and dispersions, their distribution conforms to that
of the measured values \citep[see
also][]{Fiore:2007,Coward:2013,Chadia:2013,Wang:2014}. We plot the
distribution of pseudo-redshifts in Figure (\ref{fig14}) to show
that it indeed concords with the distributions presented in
previous works
\citep{{Guetta:2005},{Bagoly:2008},{Jakobsson:2012},{Herbel:2017}}.
This important remark allows us to use the values of the
pseudo-redshifts thus obtained in statistical studies, especially when the
number of bursts is large, such as in our case. We should also
stress, however, that the pseudo-redshifts thus obtained cannot be
used for individual  bursts, due to the large uncertainties that
taint their determination.

\begin{figure}[h]
\centering
\includegraphics[angle=0, width=0.483\textwidth]{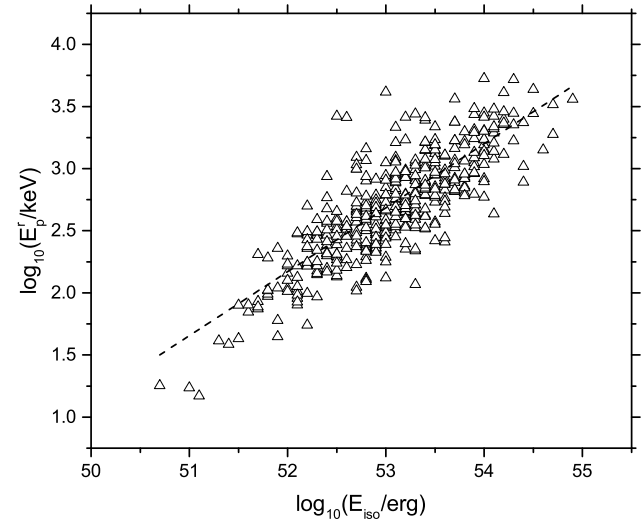}
\caption{Amati relation for 873
Fermi long GRBs. The dashed line represents a linear fit given by
the equation:
$\log{(E_p^r/keV)}=(-24.6\pm0.6)+(0.515\pm0.012)\log{(E_{iso}/erg)}$}
\label{fig15}
\end{figure}

We have also performed a validation of the results we have
obtained for peudo-redshifts by the study of the Amati relation
($\frac{E_{p,r}}{keV} = K (\frac{E_{iso}}{10^{52}~erg})^m$). This
is shown graphically in Figure (\ref{fig15}).

We obtain values for K and m that are close to those that we
obtained in our previous work, which we conducted on Swift/BAT
bursts \citep{Zitouni:2014}. In Table (\ref{tabAmati}); we also compare
our results with the original ones of \cite{Amati:2002}, which were
obtained using a sample consisting of only 9 bursts.
Many works have produced parameters for the Amati relation, e.g.
\citep{{Amati:2008MNRAS391}, {Amati:2009AA508},
{Capozziello:2010AA519A}}. Our results are in very good agreement
with all these works, particularly with regard to the slope $m$, which
is  generally given with good precision.

\begin{table}[ht]
 \renewcommand{\arraystretch}{1.5}
  \centering
\caption{$E_{iso}-E_{p,r}$ correlation. Ref.1:
\cite{Zitouni:2014}, Ref.2: \cite{Amati:2002}}
  \begin{tabular}{ccccc}
    \hline
    ~  & This work& Ref.1&   Ref.2  \\
    \hline
    K & $151^{+2}_{-5}$ & $141^{+18}_{-15}$ & $\sim 95$ \\
    m & $0.515\pm 0.012$ & $0.45\pm 0.10$  & $0.52\pm 0.06$\\
    \hline
  \end{tabular}
\label{tabAmati}
  \end{table}

\section{Intrinsic Duration Distribution}
In this section we determine the intrinsic (source frame)
durations $T^{r}_{90}$ of Fermi GRBs, using the pseudo-redshifts
obtained from the correlation relations. We study the distribution
of $\log{(T^r_{90})}$, as had been done for the BATSE, Swift, and
Fermi bursts
\citep{Zitouni:15,Shahmoradi:2015,Tarnopolski:2016c,Zhang:2016MNRAS,Kulkarni:2017}.

The distribution of the intrinsic duration is studied in the same
approach that we applied to the distribution of the observed
durations. Indeed, we use the two statistical evaluation methods,
namely the method of minimization of the $\chi^2_r$ function and
the method of the maximization of the log-likelihood function.

\begin{figure}
\begin{tabular}{c}
   \includegraphics[angle=0, width=0.45\textwidth]{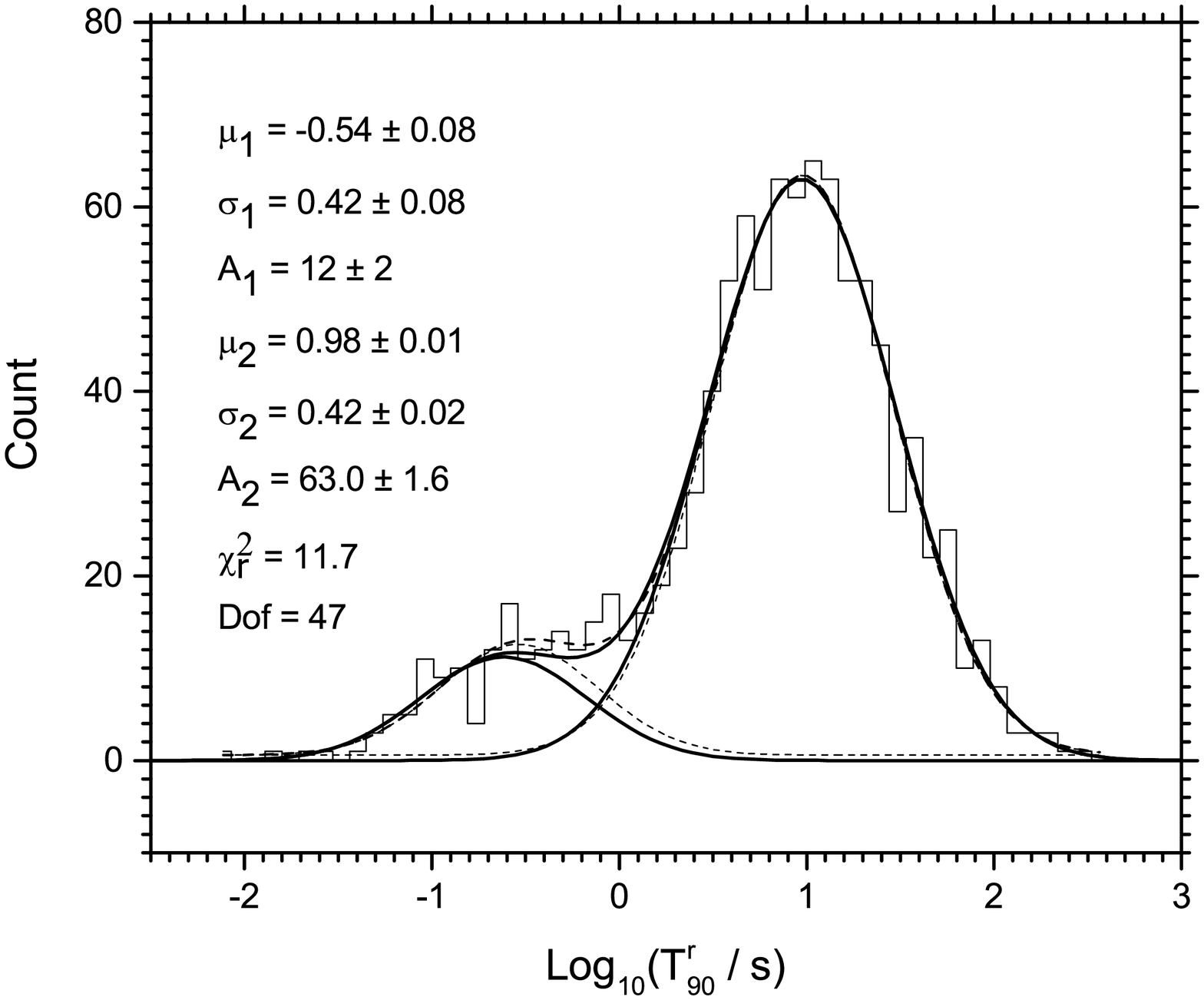} \\
   \includegraphics[angle=0, width=0.45\textwidth]{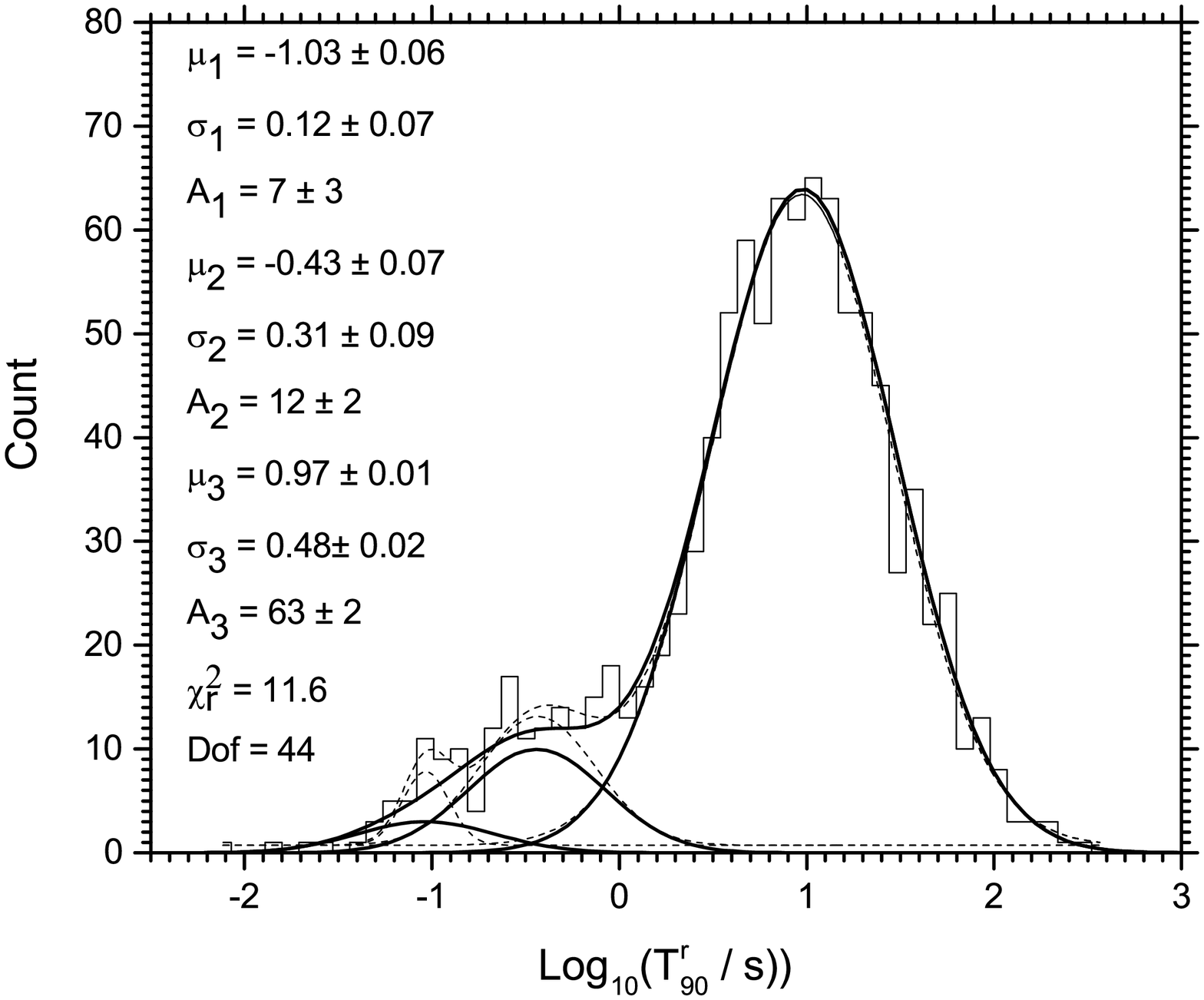}
\end{tabular}
\caption{Intrinsic duration distribution of 1017 Fermi GRBs. The
parameters A, $\mu$, and $\sigma$ characterize the Gaussian fit.
The bin width is  0.09.} \label{fig16N}
\end{figure}

The results of the log-likelihood  method are presented in Table
(\ref{tab6N}).  The (previously defined) AIC and BIC both favor
the two-group model over that of three groups. The parameters of
each model are given in the two corresponding sub-tables. Figure
(\ref{fig16N}) shows the results of the two statistical
evaluations methods: those of the $\chi^2_r$ minimization method
 are shown by dotted lines, and those of the log-likelihood method are shown by continuous lines.

The two groups/populations have the following parameters: the
durations of the peak bursts are (in the source frames):
$0.24^{+0.19}_{-0.05}~s$ and $8.7^{+1.1}_{-1.8}~s$, respectively,
and the logarithmic standard deviations are: $\sigma_1=0.44 $ and
$\sigma_2=0.50$ , respectively (Table \ref{Tab7N}). These values
are calculated from the results of the likelihood method presented
in the table (\ ref {tab6N}).

Along with results from previous works (\citet{{Zhang:2008}} and,
most recently, \citet{{Zitouni:15}}), and while we note some
differences in the centroid values of the two populations between
various works, we can state that this (and other) analysis(es) confirm(s)
the existence of two populations of bursts as {\it intrinsic} and
thus probably of inherently physical origin. The numbers of short
and long bursts recorded by each instrument may be due to the time
response of the detectors. Depending on the trigger criteria, some
gamma-ray detectors may be more responsive to one class (LGRBs)
over another (SGRBs) \citep{Hakkila:2003, Shahmoradi:2015}.

\begin{table}[ht]
\renewcommand{\arraystretch}{1.5}
  \centering
\caption{The distribution of intrinsic durations of 1017 Fermi GRBs
represented by two and three Gaussian functions (two and three groups of GRBs).
The parameters presented here are all defined in the main text.}
 \label{tab6N}
  \begin{tabular}{c|cc}
    \hline
    ~  & ~~~~~~Two groups & \\
    \hline
    i&1&2\\
    \hline
    f & $0.136\pm 0.02$ & $0.864\pm0.02$ \\
    $\mu$ & $-0.62^{+0.25}_{-0.10}$ & $0.94^{+0.05}_{-0.04}$ \\
    $\sigma$ & $0.442\pm0.079$ & $0.464\pm 0.022$\\
    $\ln{L}$&  -1055.239&\\
    BIC& 2145.101&\\
    AIC&2120.478&\\
    k&5&\\
    \hline
  \end{tabular}
  \begin{tabular}{c|ccc}
    \hline
    ~  & &Three groups&  \\
    \hline
    i&1&2&3\\
    \hline
    f & $0.03^{+0.02}_{-0.01}$& $0.10^{+0.04}_{-0.02}$&$0.87^{+0.03}_{-0.06}$\\
    $\mu$  & $-1.03^{+0.06}_{-0.07}$&$-0.44^{+0.07}_{-0.06}$&$0.98^{+0.02}_{-0.05}$\\
    $\sigma$ &$0.36\pm 0.07$&$0.37\pm0.09$&$0.497\pm0.020$\\
    $\ln{L}$&-1055.742&&\\
    BIC&2166.882&&\\
    AIC&2127.485&&\\
    k&8&&\\
    \hline
  \end{tabular}
  \end{table}

\begin{table}[ht]
\renewcommand{\arraystretch}{1.4}
  \centering
\caption{Position values of the centroids in the intrinsic
duration distributions. Ref.1: \cite{Zitouni:15}, Ref.2:
\cite{Zhang:2008}. $G_1$ and $G_2$ are the statistical parameters corresponding to the two groups/populations (SGRBs and LGRBs).}\label{Tab7N}
  \begin{tabular}{cccc}
    \hline
    & This work& Ref.1&   Ref.2  \\
    \hline
    $G_1$ & $0.24^{+0.19}_{-0.05}$ & $0.13^{+0.05}_{-0.03}$ &0.13 \\
   $ G_2$ & $8.7^{+1.1}_{-1.8}$ & $16.4^{+1.2}_{-1.1}$ & 12.30 \\
    \hline
  \end{tabular}
  \end{table}

 \section{Results and conclusion}
In this work, we have studied the distribution of the durations of
Fermi gamma-ray bursts in their source frames as well as in the
observer's frame. To obtain the intrinsic (source frame)
durations, we have assumed the validity of the correlation
relations between the luminosity $L_p$ at the spectral peak and
the energy $E_p$ at the flux peak. This assumption is needed in
order to determine the
pseudo-redshift of each burst, which then allows us to infer the
burst's duration in its source frame. However, while such
pseudo-redshifts are often tainted by substantial uncertainties,
their distribution remains close to that of the measured
redshifts. This remark, coupled with the large number of bursts
used in our study, encourages us to confidently use the
pseudo-redshifts in studying the distributions of the bursts'
durations in their source frames. It is worth mentioning that the
distributions of spectral parameters ($\alpha$,  $\beta$ and
$E_p^{Obs}$), which are used to infer peudo-redshifts from the
correlation relations between $L_p$ and $E_p$, must give the same
distribution as the measured redshifts. This is indeed the case
when we assume the validity of the correlation relations.

\begin{figure}[h]
\centering
\includegraphics[angle=0, width=0.483\textwidth]{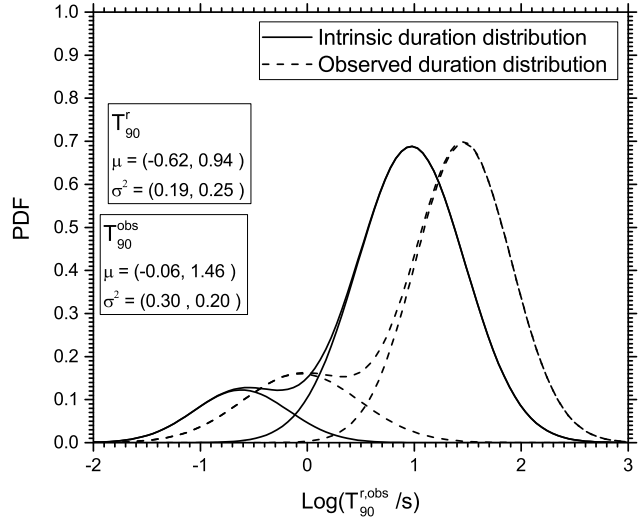}
\caption{Density of the function of distribution of observed and
intrinsic durations. The sample used here is composed of 1017
Fermi GRBS.} \label{fig18}
\end{figure}

In order to compare the distributions of the observed (measured)
and intrinsic durations, we have determined pseudo-redshifts for
1017 bursts: 144 SGRBs and 873 LGRBs. The distribution of the
observed durations of these bursts is shown in Figure (\ref{fig18})
(the dashed lines).
The fits (with Gaussian functions) were obtained by using the
maximization of the log-likelihood function $\ln{L}$; here we
adopt the two-group model for this comparison. The parameters
describing the two groups/populations are given in
 Figure (\ref{fig18}). The borderline between the two groups is
in the interval [1.7, 2.5] s. Our results are in good agreement
with the conclusions drawn by \citet{Bromberg:2013} and
\citet{Tarnopolski:2015ApSS} for Fermi bursts. They confirm the
separation between the two groups/populations around 2 seconds.

The distribution of intrinsic durations of the same sample is
shown in the same figure (\ref{fig18}) (the continuous lines).
Likewise, the fit is obtained by maximization of the
log-likelihood function. We find that two groups/populations of
bursts characterize the distribution of durations quite well and
are intrinsically separate.

We also note that the observed and intrinsic distributions are very similar,
with a shift to the right
($2.0$ s compared to $1.0$ s) for the observed durations, due
to the time dilation produced by the redshift/expansion of the
universe. This is consistent with what was found by
\citet{Tarnopolski:2016a} and \citet{Tarnopolski:2016c}. The
 shift found here may be slightly different than what
was found in those works, but that may be ascribed to
different detector characteristics (Fermi vs. Swift/BAT).

Our analysis indicates that Fermi bursts can be represented by two groups/populations. The possibility of a third one had been raised in previous works (see \citet{Zitouni:2014} and references therein) at least in some databases (Swift/BAT but not BATSE).
We find that a third component is not needed to describe the Fermi GRB durations.
Thus the issue of detector characteristics seems to be the main factor in this regard, as we had suggested (\citet{Zitouni:2014}).

Finally, the 2.0-second dividing line (for observed durations) also seems to be somewhat arbitrary and detector/instrument-dependent; for intrinsic durations, a separation around 1.0 second seems to apply, with this value being largely approximate.

 \begin{acknowledgments}
The research reported in this publication was supported by the
Mohammed Bin Rashid Space Centre (MBRSC), Dubai, UAE, under Grant
ID number 201603.SS.AUS. The authors also acknowledge the use of
the online Fermi/GBM table compiled by \citet{vonKienlin2014} and
\citet{Bhat:2016ApJS}. HZ wishes to thank the American University
of Sharjah (UAE) for hosting him for two weeks, during which part
of this work was conducted. The
authors thank the anonymous referee for very useful comments, which led to
significant improvements of the paper.
\end{acknowledgments}

\bibliographystyle{spr-mp-nameyear-cnd}
\bibliography{ref_bib_sharjah1}
\appendix
\renewcommand{\arraystretch}{1.25}
 {\scriptsize {
 \centering
\setcounter{table}{3}
\begin{longtable}{|l|l|c|c|c|c|c|c|c|c|}
\caption{Data for the 117 Fermi GRBs that were used to validate
the determination of pseudo-redshifts. 61 GRBs are used in our
calculations, the remaining 56 GRBs are eliminated because of
their spectral parameters which are found outside the chosen
intervals.  $F_{\gamma}$  and  $L_p$ are calculated in this work.}
\label{tab3}\\
  \hline
    GRB & $z_{obs}$ & $T_{90}^{obs}$ & $E_p^{obs}$ &$\alpha$&$\beta$& $\log{(F_{\gamma})}$ & $\log{(L_p)}$&$\log(R_L)$&$z_{cal}$\\
    &   & (s) & (keV) & & & ($erg\,cm^{-2}\,s^{-1}$)  & ($erg\,s^{-1}$) & &\\
  \hline
\endfirsthead

\multicolumn{10}{c}%
{{\bfseries \tablename\ \thetable{} -- continued from previous
page}} \\ \hline
    GRB & $z_{obs}$ & $T_{90}^{obs}$ & $E_p^{obs}$ &$\alpha$&$\beta$& $\log{(F_{\gamma})}$ & $\log{(L_p)}$&$\log(R_L)$&$z_{cal}$\\
    &   & (s) & (keV) & & & ($erg\,cm^{-2}\,s^{-1}$)  & ($erg\,s^{-1}$) & &\\
  \hline
\endhead

\hline \multicolumn{10}{|r|}{{Continued on next page}} \\ \hline
\endfoot

\hline \hline
\endlastfoot
GRB170705115 & 2.01 & 22.78 & 167 $\pm$ 15 & -0.694 $\pm$ 0.09 & -2.44 $\pm$ 0.25 & -5.5 & 53.03 & 0.36 &           1.14 \\
GRB170607971 & 0.557 & 22.59 & 131 $\pm$ 27 & -1.18 $\pm$ 0.13 & -2.13 $\pm$ 0.2 & -5.8 & 51.48 & -0.562 &          1.20 \\
GRB170405777 & 3.51 & 78.59 & 324 $\pm$ 41 & -0.674 $\pm$ 0.07 & -2.25 $\pm$ 0.32 & -5.5 & 53.68 & 0.268 &          2.05 \\
GRB170214649 & 2.53 & 122.88 & 455 $\pm$ 74 & -0.695 $\pm$ 0.07 & -1.96 $\pm$ 0.12 & -5.4 & 53.53 & 0.057 &         2.24 \\
GRB160821937 & 0.16 & 1.09 & 125 $\pm$ 76 & -0.663 $\pm$ 0.65 & -2.41 $\pm$ 1.11 & -6 & 49.97 & -1.167 &            0.67 \\
GRB160625945 & 1.406 & 454.67 & 594 $\pm$ 15 & -0.502 $\pm$ 0.01 & -2.05 $\pm$ 0.02 & -4.2 & 54.16 & 0.767 &        0.48 \\
GRB160509374 & 1.17 & 369.67 & 345 $\pm$ 14 & -0.752 $\pm$ 0.02 & -2.25 $\pm$ 0.06 & -4.8 & 53.21 & 0.267 &         0.82 \\
GRB151111356 & 3.5 & 46.34 & 108 $\pm$ 44 & 0.311 $\pm$ 1.15 & -2.77 $\pm$ 1.78 & -6.9 & 52.21 & -0.437 &           11.06 \\
GRB151027166 & 0.81 & 123.39 & 224 $\pm$ 46 & -0.855 $\pm$ 0.11 & -2.05 $\pm$ 0.17 & -5.8 & 51.91 & -0.608 &        2.18 \\
GRB150821406 & 0.755 & 103.43 & 291 $\pm$ 52 & -0.843 $\pm$ 0.09 & -2.13 $\pm$ 0.21 & -5.7 & 51.89 & -0.787 &       2.76 \\
GRB150514774 & 0.807 & 10.81 & 55 $\pm$ 4 & -0.765 $\pm$ 0.15 & -2.77 $\pm$ 0.19 & -5.9 & 51.66 & 0.12 &            0.71 \\
GRB150403913 & 2.06 & 22.27 & 509 $\pm$ 34 & -0.672 $\pm$ 0.03 & -2.19 $\pm$ 0.08 & -5 & 53.61 & 0.155 &            1.57 \\
GRB150314205 & 1.758 & 10.69 & 298 $\pm$ 9 & -0.406 $\pm$ 0.03 & -2.54 $\pm$ 0.08 & -4.8 & 53.62 & 0.61 &           0.73 \\
GRB141121414 & 1.47 & 3.84 & 144 $\pm$ 36 & 0.538 $\pm$ 0.71 & -2.4 $\pm$ 0.66 & -6.4 & 51.78 & -0.65 &             5.44 \\
GRB141004973 & 0.573 & 2.56 & 31 $\pm$ 10 & 0.43 $\pm$ 1.47 & -1.88 $\pm$ 0.07 & -6 & 51.31 & 0.25 &                0.42 \\
GRB140907672 & 1.21 & 35.84 & 101 $\pm$ 20 & -0.511 $\pm$ 0.32 & -2.7 $\pm$ 0.73 & -6.4 & 51.57 & -0.534 &          3.01 \\
GRB140817293 & 0.018 & 16.13 & 132 $\pm$ 21 & -0.464 $\pm$ 0.24 & -2.67 $\pm$ 0.45 & -5.9 & 47.99 & -3.763 &        1.82 \\
GRB140808038 & 3.29 & 4.48 & 144 $\pm$ 14 & -0.107 $\pm$ 0.17 & -3.01 $\pm$ 0.62 & -5.9 & 53.07 & 0.262 &           2.05 \\
GRB140801792 & 1.32 & 7.17 & 122 $\pm$ 4 & 0.045 $\pm$ 0.08 & -4.05 $\pm$ 0.8 & -5.6 & 52.44 & 0.167 &              1.04 \\
GRB140703026 & 3.14 & 83.97 & 232 $\pm$ 69 & -0.699 $\pm$ 0.21 & -2.34 $\pm$ 0.66 & -6.2 & 52.85 & -0.273 &         6.31 \\
GRB140508128 & 1.027 & 44.29 & 369 $\pm$ 20 & -0.55 $\pm$ 0.04 & -2.37 $\pm$ 0.09 & -4.9 & 52.98 & 0.034 &          0.98 \\
GRB140506880 & 0.889 & 64.13 & 138 $\pm$ 21 & -0.484 $\pm$ 0.21 & -2.43 $\pm$ 0.25 & -5.7 & 51.99 & -0.218 &        1.24 \\
GRB140423356 & 3.26 & 95.23 & 176 $\pm$ 86 & -0.486 $\pm$ 0.46 & -2.06 $\pm$ 0.51 & -6.3 & 52.74 & -0.214 &         5.77 \\
GRB140304557 & 5.283 & 31.23 & 165 $\pm$ 92 & -0.56 $\pm$ 0.53 & -2.06 $\pm$ 0.57 & -6.4 & 53.1 & -0.075 &          6.69 \\
GRB140213807 & 1.208 & 18.62 & 84 $\pm$ 3 & -0.844 $\pm$ 0.06 & -2.89 $\pm$ 0.17 & -5.5 & 52.44 & 0.461 &           0.65 \\
GRB140206304 & 2.73 & 27.26 & 112 $\pm$ 7 & 0.641 $\pm$ 0.21 & -2.5 $\pm$ 0.12 & -5.6 & 53.26 & 0.718 &             0.90 \\
GRB131231198 & 0.644 & 31.23 & 313 $\pm$ 11 & -0.792 $\pm$ 0.02 & -2.66 $\pm$ 0.11 & -4.8 & 52.52 & -0.17 &         0.82 \\
GRB131108862 & 2.4 & 18.18 & 333 $\pm$ 35 & -0.598 $\pm$ 0.07 & -1.94 $\pm$ 0.07 & -5.3 & 53.48 & 0.245 &           1.53 \\
GRB131011741 & 1.874 & 77.06 & 168 $\pm$ 88 & -0.472 $\pm$ 0.43 & -1.63 $\pm$ 0.12 & -6.1 & 52.59 & -0.054 &        2.11 \\
GRB130518580 & 2.488 & 48.58 & 467 $\pm$ 31 & -0.75 $\pm$ 0.03 & -2.15 $\pm$ 0.07 & -4.9 & 53.91 & 0.424 &          1.20 \\
GRB130427324 & 0.34 & 138.24 & 718 $\pm$ 6 & -0.454 $\pm$ 0.01 & -3.13 $\pm$ 0.04 & -3.7 & 53.05 & -0.072 &         0.38 \\
GRB121031949 & 0.113 & 242.44 & 300 $\pm$ 98 & -0.693 $\pm$ 0.21 & -2.74 $\pm$ 1.97 & -6.1 & 49.48 & -2.905 &       10.12 \\
GRB120922939 & 3.1 & 182.28 & 65 $\pm$ 24 & -1.047 $\pm$ 0.62 & -2.48 $\pm$ 0.61 & -6.7 & 52.3 & 0.07 &             2.76 \\
GRB120716712 & 2.48 & 237.06 & 106 $\pm$ 17 & -0.389 $\pm$ 0.22 & -2.14 $\pm$ 0.17 & -6 & 52.83 & 0.374 &           1.31 \\
GRB120119170 & 1.728 & 55.3 & 274 $\pm$ 32 & -0.842 $\pm$ 0.06 & -2.35 $\pm$ 0.24 & -5.5 & 52.88 & -0.066 &         1.93 \\
GRB111228657 & 0.716 & 99.84 & 95 $\pm$ 12 & -1.374 $\pm$ 0.08 & -2.73 $\pm$ 0.51 & -5.9 & 51.6 & -0.289 &          1.07 \\
GRB111107035 & 2.893 & 12.03 & 129 $\pm$ 76 & -0.649 $\pm$ 0.73 & -2.55 $\pm$ 1.63 & -6.6 & 52.31 & -0.36 &         6.69 \\
GRB110721200 & 0.382 & 21.82 & 225 $\pm$ 18 & -0.64 $\pm$ 0.05 & -1.93 $\pm$ 0.04 & -5.2 & 51.73 & -0.6 &           0.84 \\
GRB110213220 & 1.46 & 34.3 & 81 $\pm$ 10 & -1.059 $\pm$ 0.13 & -2.67 $\pm$ 0.37 & -5.9 & 52.34 & 0.311 &            0.95 \\
GRB101219686 & 0.552 & 51.01 & 87 $\pm$ 23 & 0.496 $\pm$ 0.98 & -2.64 $\pm$ 0.85 & -6.6 & 50.5 & -1.261 &           4.05 \\
GRB100906576 & 1.727 & 110.59 & 146 $\pm$ 25 & -0.603 $\pm$ 0.16 & -2.14 $\pm$ 0.16 & -5.7 & 52.78 & 0.268 &        1.15 \\
GRB100814160 & 1.44 & 150.53 & 128 $\pm$ 29 & 0.911 $\pm$ 0.72 & -1.84 $\pm$ 0.14 & -6 & 52.33 & -0.007 &           1.48 \\
GRB100728439 & 2.106 & 10.24 & 63 $\pm$ 21 & -0.178 $\pm$ 0.62 & -1.84 $\pm$ 0.12 & -6.1 & 52.55 & 0.535 &          0.87 \\
GRB100728095 & 1.567 & 165.38 & 446 $\pm$ 56 & -0.454 $\pm$ 0.09 & -2.47 $\pm$ 0.38 & -5.5 & 52.81 & -0.429 &       3.60 \\
GRB100724029 & 1.288 & 114.69 & 472 $\pm$ 39 & -0.67 $\pm$ 0.04 & -2 $\pm$ 0.07 & -5.2 & 53.02 & -0.179 &           1.77 \\
GRB100615083 & 1.398 & 37.38 & 48 $\pm$ 7 & 0.311 $\pm$ 0.5 & -2.12 $\pm$ 0.1 & -6.1 & 52.09 & 0.445 &              0.75 \\
GRB100414097 & 1.368 & 26.5 & 478 $\pm$ 40 & -0.724 $\pm$ 0.04 & -2.52 $\pm$ 0.26 & -5.3 & 52.92 & -0.318 &         2.38 \\
GRB100216422 & 0.038 & 0.19 & 510 $\pm$ 382 & 0.036 $\pm$ 0.82 & -1.72 $\pm$ 0.28 & -5.6 & 49.4 & -2.624 &          1.01 \\
GRB100206563 & 0.41 & 0.18 & 566 $\pm$ 117 & -0.222 $\pm$ 0.18 & -2.38 $\pm$ 0.45 & -5 & 51.97 & -0.336 &           0.63 \\
GRB091127976 & 0.49 & 8.7 & 55 $\pm$ 2 & -0.509 $\pm$ 0.08 & -2.27 $\pm$ 0.03 & -5.3 & 51.81 & 0.4 &                0.31 \\
GRB091020900 & 1.71 & 24.26 & 240 $\pm$ 143 & -0.912 $\pm$ 0.26 & -1.68 $\pm$ 0.09 & -5.9 & 52.65 & -0.202 &        2.68 \\
GRB091003191 & 0.897 & 20.22 & 390 $\pm$ 25 & -0.596 $\pm$ 0.04 & -2.33 $\pm$ 0.11 & -5.1 & 52.69 & -0.252 &        1.32 \\
GRB090926181 & 2.106 & 13.76 & 349 $\pm$ 9 & -0.464 $\pm$ 0.02 & -2.7 $\pm$ 0.1 & -4.7 & 53.87 & 0.664 &            0.77 \\
GRB090618353 & 0.54 & 112.39 & 426 $\pm$ 24 & -0.958 $\pm$ 0.03 & -2.87 $\pm$ 0.26 & -4.9 & 52.29 & -0.56 &         1.17 \\
GRB090516353 & 4.109 & 123.07 & 70 $\pm$ 50 & -0.555 $\pm$ 0.92 & -1.79 $\pm$ 0.15 & -6.3 & 53.05 & 0.621 &         1.16 \\
GRB090424592 & 0.544 & 14.14 & 187 $\pm$ 6 & -0.843 $\pm$ 0.02 & -2.94 $\pm$ 0.16 & -5 & 52.17 & -0.119 &           0.65 \\
GRB081222204 & 2.77 & 18.88 & 167 $\pm$ 19 & -0.735 $\pm$ 0.09 & -2.59 $\pm$ 0.38 & -5.8 & 53.08 & 0.257 &          1.77 \\
GRB081121858 & 2.512 & 41.98 & 192 $\pm$ 66 & -0.195 $\pm$ 0.44 & -1.76 $\pm$ 0.1 & -5.8 & 53.08 & 0.209 &          1.67 \\
GRB081025349 & 0.39 & 22.53 & 406 $\pm$ 122 & -0.415 $\pm$ 0.22 & -2.77 $\pm$ 2.21 & -6 & 50.87 & -1.875 &          0.02 \\
GRB081028538 & 3.038 & 13.31 & 64 $\pm$ 9 & 0.171 $\pm$ 0.45 & -2.48 $\pm$ 0.24 & -6.2 & 52.7 & 0.49 &              1.32 \\
GRB080804972 & 2.205 & 24.7 & 163 $\pm$ 32 & -0.189 $\pm$ 0.28 & -2.8 $\pm$ 1.14 & -6.3 & 52.35 & -0.342 &          4.42 \\
GRB170817529 & 0.01 & 2.05 & 177 $\pm$ 99 & 1.728 $\pm$ 2.99 & -2.47 $\pm$ 1.47 & ~~ & ~~ & ~~ &                    ~~ \\
GRB170714049 & 0.793 & 0.22 & 487 $\pm$ 127 & 0.539 $\pm$ 0.68 & -17.62 $\pm$ 5840000 & ~~ & ~~ & ~~ &              ~~ \\
GRB170428136 & 0.454 & 30.46 & 131 $\pm$ 42 & -0.568 $\pm$ 0.39 & -7.53 $\pm$ 1366.51 & ~~ & ~~ & ~~ &              ~~ \\
GRB170113420 & 1.968 & 49.15 & 235 $\pm$ 574 & -1.354 $\pm$ 0.71 & -1.96 $\pm$ 0.72 & ~~ & ~~ & ~~ &                ~~ \\
GRB161129300 & 0.645 & 36.1 & 276 $\pm$ 59 & -0.878 $\pm$ 0.14 & -13.78 $\pm$ 1630000 & ~~ & ~~ & ~~ &              ~~ \\
GRB161117066 & 1.549 & 122.18 & 71 $\pm$ 7 & -0.813 $\pm$ 0.18 & 4.06 $\pm$ 29.21 & ~~ & ~~ & ~~ &                  ~~ \\
GRB161017745 & 2.013 & 32.26 & 417 $\pm$ 138 & -1.23 $\pm$ 0.14 & -5.1 $\pm$ 123.83 & ~~ & ~~ & ~~ &                ~~ \\
GRB161014522 & 2.823 & 36.61 & 202 $\pm$ 27 & -0.518 $\pm$ 0.13 & -11.96 $\pm$ 154897.1 & ~~ & ~~ & ~~ &            ~~ \\
GRB161001045 & 0.891 & 2.24 & 442 $\pm$ 57 & -0.726 $\pm$ 0.09 & -13.31 $\pm$ 349696 & ~~ & ~~ & ~~ &               ~~ \\
GRB160804065 & 0.736 & 131.59 & 87 $\pm$ 59 & -1.019 $\pm$ 0.42 & -2.03 $\pm$ 3.25 & ~~ & ~~ & ~~ &                 ~~ \\
GRB160624477 & 0.483 & 0.45 & 678 $\pm$ 258 & 0.115 $\pm$ 0.59 & -19.11 $\pm$ 39570000 & ~~ & ~~ & ~~ &             ~~ \\
GRB160314473 & 0.727 & 1.66 & 709 $\pm$ 989 & -1.181 $\pm$ 0.33 & -5.12 $\pm$ 226.2 & ~~ & ~~ & ~~ &                ~~ \\
GRB160228034 & 1.64 & 16.13 & 16 $\pm$ 119 & -0.589 $\pm$ 35.32 & -1.9 $\pm$ 0.2 & ~~ & ~~ & ~~ &                   ~~ \\
GRB160227831 & 2.38 & 7.68 & 511 $\pm$ 36 & -0.347 $\pm$ 0.07 & -18.63 $\pm$ 21200000 & ~~ & ~~ & ~~ &              ~~ \\
GRB150727793 & 0.313 & 49.41 & 86 $\pm$ 54 & 1.732 $\pm$ 3.6 & -1.7 $\pm$ 0.19 & ~~ & ~~ & ~~ &                     ~~ \\
GRB150512432 & 0.104 & 123.14 & 219 $\pm$ 36 & -0.776 $\pm$ 0.15 & -11.29 $\pm$ 84293.06 & ~~ & ~~ & ~~ &           ~~ \\
GRB150424403 & 0.3 & 36.1 & 148 $\pm$ 53 & -1.049 $\pm$ 0.25 & -3.49 $\pm$ 8.65 & ~~ & ~~ & ~~ &                    ~~ \\
GRB150301818 & 1.517 & 13.31 & 174 $\pm$ 37 & -0.952 $\pm$ 0.17 & -8.47 $\pm$ 5166.64 & ~~ & ~~ & ~~ &              ~~ \\
GRB150101641 & 0.134 & 0.08 & 23 $\pm$ 5 & 1.161 $\pm$ 2.95 & -1.6 $\pm$ 1.08 & ~~ & ~~ & ~~ &                      ~~ \\
GRB141225959 & 0.915 & 56.32 & 145 $\pm$ 47 & 1.021 $\pm$ 1.12 & -1.95 $\pm$ 0.27 & ~~ & ~~ & ~~ &                  ~~ \\
GRB141221338 & 1.452 & 23.81 & 321 $\pm$ 114 & -1.075 $\pm$ 0.15 & -6.32 $\pm$ 663.11 & ~~ & ~~ & ~~ &              ~~ \\
GRB141220252 & 1.32 & 7.62 & 215 $\pm$ 13 & -0.525 $\pm$ 0.07 & -6.88 $\pm$ 114.6 & ~~ & ~~ & ~~ &                  ~~ \\
GRB141118678 & 0.108 & 4.35 & 317 $\pm$ 41 & -0.536 $\pm$ 0.12 & -13.97 $\pm$ 789873.6 & ~~ & ~~ & ~~ &             ~~ \\
GRB140606133 & 0.384 & 22.78 & 881 $\pm$ 350 & -1.268 $\pm$ 0.06 & 0.28 $\pm$ 5.54 & ~~ & ~~ & ~~ &                 ~~ \\
GRB140512814 & 0.725 & 147.97 & 575 $\pm$ 95 & -1 $\pm$ 0.06 & -4.37 $\pm$ 15.75 & ~~ & ~~ & ~~ &                   ~~ \\
GRB140311885 & 4.954 & 72.19 & 353 $\pm$ 290 & -1.138 $\pm$ 0.3 & -6.96 $\pm$ 4223.74 & ~~ & ~~ & ~~ &              ~~ \\
GRB140219824 & 0.12 & 77.06 & 32 $\pm$ 19 & 1.534 $\pm$ 3.98 & -1.76 $\pm$ 0.15 & ~~ & ~~ & ~~ &                    ~~ \\
GRB131105087 & 1.686 & 112.64 & 453 $\pm$ 76 & -1.032 $\pm$ 0.06 & -16.38 $\pm$ 14810000 & ~~ & ~~ & ~~ &           ~~ \\
GRB131004904 & 0.717 & 1.15 & 144 $\pm$ 90 & -1.254 $\pm$ 0.46 & -19.41 $\pm$ 3.146e+09 & ~~ & ~~ & ~~ &            ~~ \\
GRB130925173 & 0.347 & 215.56 & 94 $\pm$ 12 & -1.305 $\pm$ 0.12 & -6.4 $\pm$ 270.65 & ~~ & ~~ & ~~ &                ~~ \\
GRB130702004 & 0.145 & 58.88 & 13 $\pm$ 54 & 2.078 $\pm$ 85.25 & -1.75 $\pm$ 0.05 & ~~ & ~~ & ~~ &                  ~~ \\
GRB130612141 & 2.006 & 7.42 & 23 $\pm$ 7 & 4.747 $\pm$ 7.92 & -2.15 $\pm$ 0.18 & ~~ & ~~ & ~~ &                     ~~ \\
GRB130610133 & 2.092 & 21.76 & 150 $\pm$ 56 & -1.089 $\pm$ 0.3 & -18 $\pm$ 679300000 & ~~ & ~~ & ~~ &               ~~ \\
GRB130420313 & 1.297 & 104.96 & 59 $\pm$ 6 & -0.03 $\pm$ 0.58 & -21.3 $\pm$ 425600000 & ~~ & ~~ & ~~ &              ~~ \\
GRB121211574 & 1.023 & 5.63 & 111 $\pm$ 45 & -0.645 $\pm$ 0.54 & -3 $\pm$ 3.08 & ~~ & ~~ & ~~ &                     ~~ \\
GRB121128212 & 2.2 & 17.34 & 115 $\pm$ 7 & -0.459 $\pm$ 0.11 & -15.64 $\pm$ 2160000 & ~~ & ~~ & ~~ &                ~~ \\
GRB121123421 & 2.7 & 102.34 & 164 $\pm$ 26 & -0.384 $\pm$ 0.26 & -8.77 $\pm$ 2722.15 & ~~ & ~~ & ~~ &               ~~ \\
GRB120909070 & 3.93 & 112.07 & 705 $\pm$ 361 & -0.912 $\pm$ 0.16 & -2.75 $\pm$ 3.13 & ~~ & ~~ & ~~ &                ~~ \\
GRB120907017 & 0.97 & 5.76 & 134 $\pm$ 43 & -0.936 $\pm$ 0.3 & -8.88 $\pm$ 12518.22 & ~~ & ~~ & ~~ &                ~~ \\
GRB120811649 & 2.671 & 14.34 & 33 $\pm$ 6 & 4.591 $\pm$ 4.37 & -2.23 $\pm$ 0.16 & ~~ & ~~ & ~~ &                    ~~ \\
GRB120729456 & 0.8 & 25.47 & 27 $\pm$ 234 & -0.053 $\pm$ 38.57 & -1.63 $\pm$ 0.08 & ~~ & ~~ & ~~ &                  ~~ \\
GRB120118709 & 2.943 & 37.83 & 67 $\pm$ 13 & -0.932 $\pm$ 0.35 & -5.94 $\pm$ 145.97 & ~~ & ~~ & ~~ &                ~~ \\
GRB110818860 & 3.36 & 67.07 & 55 $\pm$ 34 & 1.246 $\pm$ 2.92 & -1.64 $\pm$ 0.12 & ~~ & ~~ & ~~ &                    ~~ \\
GRB110128073 & 2.339 & 12.16 & 323 $\pm$ 207 & -0.899 $\pm$ 0.34 & -13.28 $\pm$ 2544000 & ~~ & ~~ & ~~ &            ~~ \\
GRB110106893 & 0.618 & 35.52 & 354 $\pm$ 306 & -1.187 $\pm$ 0.28 & -7.38 $\pm$ 7202.92 & ~~ & ~~ & ~~ &             ~~ \\
GRB100816026 & 0.805 & 2.04 & 132 $\pm$ 24 & -0.13 $\pm$ 0.33 & 2.17 $\pm$ 1.39 & ~~ & ~~ & ~~ &                    ~~ \\
GRB091024372 & 1.092 & 93.95 & 652 $\pm$ 972 & -1.209 $\pm$ 0.42 & -2.66 $\pm$ 4.39 & ~~ & ~~ & ~~ &                ~~ \\
GRB090927422 & 1.37 & 0.51 & 96 $\pm$ 15 & 2 $\pm$ 1.56 & -10.93 $\pm$ 2663.18 & ~~ & ~~ & ~~ &                     ~~ \\
GRB090926914 & 1.24 & 55.55 & 111 $\pm$ 14 & -0.121 $\pm$ 0.32 & -6.25 $\pm$ 76.75 & ~~ & ~~ & ~~ &                 ~~ \\
GRB090423330 & 8 & 7.17 & 80 $\pm$ 21 & -0.616 $\pm$ 0.54 & -7.86 $\pm$ 1923.03 & ~~ & ~~ & ~~ &                    ~~ \\
GRB090328401 & 0.736 & 61.7 & 458 $\pm$ 35 & -0.784 $\pm$ 0.04 & -4.45 $\pm$ 7.14 & ~~ & ~~ & ~~ &                  ~~ \\
GRB090102122 & 1.547 & 26.62 & 378 $\pm$ 23 & -0.274 $\pm$ 0.07 & -7.32 $\pm$ 157.82 & ~~ & ~~ & ~~ &               ~~ \\
GRB081008832 & 1.969 & 150.01 & 173 $\pm$ 49 & -0.662 $\pm$ 0.3 & -12.3 $\pm$ 434886.5 & ~~ & ~~ & ~~ &             ~~ \\
GRB080928628 & 1.69 & 14.34 & 83 $\pm$ 33 & -1.139 $\pm$ 0.45 & -2.92 $\pm$ 2.52 & ~~ & ~~ & ~~ &                   ~~ \\
GRB080916406 & 0.689 & 46.34 & 241 $\pm$ 37 & -0.428 $\pm$ 0.17 & -4.83 $\pm$ 20.44 & ~~ & ~~ & ~~ &                ~~ \\
GRB080810549 & 3.35 & 107.46 & 735 $\pm$ 489 & -1.18 $\pm$ 0.14 & -2.43 $\pm$ 1.75 & ~~ & ~~ & ~~ &                 ~~ \\
\end{longtable}
} }
\newpage
\end{document}